\documentclass{aa}
\pdfoutput=1
\usepackage{natbib}
\usepackage{graphicx}
\usepackage{url}
\usepackage{txfonts}

\bibpunct{(}{)}{;}{a}{}{,} % to follow the A&A style

\usepackage{color}

\graphicspath{{./figs/}}

\begin{document}

\title{Long term variability of Cygnus X-1:} 
\subtitle{VII. Orbital variability of the focussed wind in Cyg X-1 /
  HDE 226868 system}

\titlerunning{Orbital variability of the focussed wind in Cygnus X-1 /
  HDE 226868 system}

\author{\mbox{V.~Grinberg\inst{\ref{affil:mit}}} \and
\mbox{M.A.~Leutenegger\inst{\ref{affil:cresst},\ref{affil:gsfc}}} \and 
\mbox{N.~Hell\inst{\ref{affil:remeis},\ref{affil:llnl}}} \and
\mbox{K.~Pottschmidt\inst{\ref{affil:cresst},\ref{affil:gsfc}}} \and 
\mbox{M.~B\"ock\inst{\ref{affil:mpifr},\ref{affil:remeis}}} \and 
\mbox{J.A.~Garc\'{i}a\inst{\ref{affil:cfa}}} \and
\mbox{M.~Hanke\inst{\ref{affil:remeis}}} \and 
\mbox{M.A.~Nowak\inst{\ref{affil:mit}}} \and
\mbox{J.O.~Sundqvist\inst{\ref{affil:delaware},\ref{affil:lmu}}} \and
\mbox{R.H.D.~Townsend\inst{\ref{affil:wisconsin}}} \and
\mbox{J.~Wilms\inst{\ref{affil:remeis}}}
} \offprints{V.~Grinberg,\\ e-mail: {grinberg@space.mit.edu}}
\institute{ Massachusetts Institute of Technology, Kavli Institute for
  Astrophysics and Space Research, Cambridge, MA 02139, USA
 \label{affil:mit}
\and 
CRESST, University of Maryland Baltimore County, 1000 Hilltop Circle,
Baltimore, MD 21250, USA \label{affil:cresst}
\and 
NASA Goddard Space Flight Center, Astrophysics Science Division, Greenbelt, MD 20771, USA \label{affil:gsfc}
\and
Dr.\ Karl-Remeis-Sternwarte  and ECAP, FAU
Erlangen-N\"urnberg, Sternwartstr.~7, 96049 Bamberg, Germany
\label{affil:remeis}
\and 
Lawrence Livermore National Laboratory, 7000 East Ave.,
Livermore, CA 94550, USA \label{affil:llnl}
\and
Max-Planck-Institut f\"ur Radioastronomie, Auf dem H\"ugel 69, 53121 
Bonn, Germany \label{affil:mpifr}
\and
Harvard-Smithsonian Center for Astrophysics, 60 Garden Street, 
Cambridge, MA 02138, USA \label{affil:cfa}
\and
University of Delaware, Bartol Research Institute, Newark, DE 19716, 
USA \label{affil:delaware}
\and
Institut f\"ur Astronomie und Astrophysik der Universit\"at M\"unchen, 
Scheinerstr.~1, 81679 M\"unchen, Germany \label{affil:lmu}
\and
Department of Astronomy, University of Wisconsin-Madison, 2535 
Sterling Hall, 475 North Charter Street, Madison, WI 53706, USA
\label{affil:wisconsin}
}
\date {Received: --- / Accepted: ---}

\abstract{Binary systems with an accreting compact object are a unique
  chance to investigate the strong, clumpy, line-driven winds of early
  type supergiants by using the compact object's X-rays to probe
  the wind structure. We analyze the two-component wind of HDE~226868,
  the O9.7Iab giant companion of the black hole Cyg~X-1 using 4.77\,Ms
  of RXTE observations of the system taken over the course of 16
  years. Absorption changes strongly over the 5.6\,d binary orbit, but
  also shows a large scatter at a given orbital phase, especially at
  superior conjunction. The orbital variability is most prominent when
  the black hole is in the hard X-ray state. Our data are poorer for
  the intermediate and soft state, but show signs for orbital
  variability of the absorption column in the intermediate state. We
  quantitatively compare the data in the hard state to a toy model of
  a focussed Castor-Abbott-Klein-wind: as it does not incorporate
  clumping, the model does not describe the observations well. A
  qualitative comparison to a simplified simulation of clumpy winds
  with spherical clumps shows good agreement in the distribution of
  the equivalent hydrogen column density for models with a porosity
  length on the order of the stellar radius at inferior conjunction;
  we conjecture that the deviations between data and model at superior
  conjunction could be either due to lack of a focussed wind component
  in the model or a more complicated clump structure.}


\maketitle

\section{Introduction}\label{sect:intro}

Early type supergiants show strong line driven winds (CAK mechanism;
\citealt*{Castor_1975a}; \citealt{Morton_1967a}; \citealt{Lucy_1970a})
and typical mass loss rates of a few $10^{-6} M_{\odot} / \mathrm{yr}$
\citep[e.g.,][]{Puls_2006a}. These high velocity winds that can reach
terminal velocities $\varv_{\infty} > 2000\,\mathrm{km}\,\mathrm{s}^{-1}$
\citep[e.g.,][]{Muijres_2012a} are perturbed and clumpy
\citep{Owocki_1988a, Feldmeier_1997a, Puls_2006a,Oskinova_2012a,
  Sundqvist_2013a} with over 90\% of the wind mass concentrated in
less than $\sim$10\% of the wind volume \citep[][for Vela~X-1 and
Cyg~X-1, respectively]{Sako_1999a,Rahoui_2011a}. Binary systems
consisting of an O/B type giant and a compact object offer us the
unique chance to investigate these winds by using the X-ray source as
a test probe.

\object{Cygnus X-1} is in such a high mass X-ray binary with the O9.7Iab
supergiant \object{HDE 226868} \citep{Walborn_1973a}. Located $\sim$1.86\,kpc
away \citep[][see also \citealt{Xiang_2011a}]{Reid_2011a}, it is one
of the brightest and best-observed black hole binaries. The orbital
period is $P=5.599829 \pm 0.000016$\,d 
with $T_0 = 52872.288 \pm 0.009$ \citep{Gies_2008a}. The source also
shows a superorbital period that has been reported as $\sim$300\,d
\citep{Priedhorsky_1983a}, then as $\sim$150\,d
\citep{Brocksopp_1999b,Benlloch_2004a,Ibragimov_2007a}, and lately as
having changed back to $\sim$300\,d \citep{Zdziarski_2011a}.

The intrinsic variability of Cyg X-1 on timescales of hours
\citep{Boeck_2011a} to years \citep{Pottschmidt_2003b, Wilms_2006a,
  Shaposhnikov_2006a, Axelsson_2006a,Grinberg_2013a} can be classified
in terms of X-ray spectral states that also show distinct timing
characteristics on timescales below 1\,s and are similar for most
black hole binaries \citep[for a review see,
e.g.,][]{Belloni_2010a}. In particular, the hard state spectrum above
$\sim$2\,keV is dominated by a power law component with photon index
$\Gamma \sim 1.7$, while in the soft state the thermal emission from
the accretion disk is prominent and the contribution from a steeper
power law component is low. 

\citet{Orosz_2011a} determine the inclination of the binary system to
$i=27\fdg{}1\pm0\fdg{}8$, the black hole mass to $M_{\mathrm{BH}} =
(14.8\pm 1.0) M_{\odot}$, and the companion mass to
$M_{\mathrm{\star}} = (19.2\pm 1.9) M_{\odot}$, the latter value being
questioned by \citet{Ziolkowski_2014a}, who estimate a range of
25--35\,$M_{\odot}$ with a most likely value of 27\,$M_{\odot}$. As
HDE~226868 fills $\gtrsim$90\% of its Roche lobe \citep{Gies_2003a}
and has a strong \citep[a few
$10^{-6}$\,$M_{\odot}\,\mathrm{yr}^{-1}$,][]{Gies_2003a} fast wind,
the black hole accretes via a focussed wind \citep[][and
others]{Friend_1982a,Gies_1986a,Gies_1986b,Sowers_1998a,Miller_2005a,Hanke_2009a,Hell_2013a,Miskovicova_2014a}.
Values for $\varv_{\infty}$ for HDE~226868 vary in the
  literature: \citet{Davis_1983a} obtain $(2300 \pm 400)
  \,\mathrm{km}\,\mathrm{s}^{-1}$ in the hard state, but
  \citet{Vrtilek_2008a} $1420\,\mathrm{km}\,\mathrm{s}^{-1}$ in the
  soft state. \citet{Herrero_1995a} adopt
  $2100\,\mathrm{km}\,\mathrm{s}^{-1}$. \citet{Gies_2008a} obtain
  $1200\,\mathrm{km}\,\mathrm{s}^{-1}$ during a soft state, but list
  $1600\,\mathrm{km}\,\mathrm{s}^{-1}$ as more likely.

The inclination of the system implies that our line of sight probes
different regions of the wind at different orbital phases. Orbital
variability of the absorption column density in the hard state has
been reported by \citet{Feng_2002a}, \citet{Ibragimov_2005a}, and
\citet{Boroson_2010a}. Dips in the (soft) X-ray lightcurves of Cyg X-1
are interpreted as signatures of clumps in the stellar wind. They
occur preferably at superior conjunction ($\phi_{\mathrm{orb}} \sim
0$), i.e., when the O-star companion is between the observer and the
black hole
\citep[e.g.,][]{Li_1974a,Mason_1974a,Parsignault_1976a,Pravdo_1980a,Remillard_1984a,Kitamoto_1984a,Balucinska-Church_2000a,
  Poutanen_2008a,Boroson_2010a}.

The 16\,years of mainly bi-weekly Cyg X-1 observations (1996--2011)
with RXTE's Proportional Counter Array \citep[PCA;][]{Jahoda_2006a}
and High Energy X-ray Timing Experiment
\citep[HEXTE;][]{Rothschild_1998a} allowed us to study the long-term
spectral and X-ray timing evolution of the source in unprecedented
detail in the previous paper of the series
\citep{Pottschmidt_2003b,Gleissner_2004a,Gleissner_2004b,Wilms_2006a,Grinberg_2013a,Grinberg_2014a}.
In this paper, we use these data to analyze the orbital variability of
absorption due to the focussed wind. The exceptional exposure reveals
subtle effects such as the orbital variability in intermediate states
and rare events with very high absorption.

In Sect.~\ref{sect:data}, we introduce the data used and the spectral
models applied to describe them. In Sect.~\ref{sect:orbvar}, we
discuss the observed orbital variability of absorption in hard,
intermediate, and soft states. In Sect.~\ref{sect:disc}, we compare
our measurements with previous results and with different wind models,
namely with a toy model for a CAK-wind and with simulations of clumpy
winds in single stars. We summarize our results in
Sect.~\ref{sect:sum}.

\section{Data and spectral analysis}\label{sect:data}

\citet{Ibragimov_2005a} used 42 Ginga-OSSE and RXTE-OSSE observations
of Cyg X-1 taken mainly in the hard state to show orbital variability
of absorption by demonstrating an increased equivalent hydrogen column
density, $N_{\mathrm{H}}$, around $\phi_{\mathrm{orb}} \sim 0$. In
\citet{Wilms_2006a}, we analyzed 202 observations from 1999--2004
using RXTE-PCA and, where available, -HEXTE spectra with models
similar to the one presented below and with a time resolution of
individual RXTE observations that were typically of several satellite
orbits in length, i.e., had exposure times of
5--10\,ks. \citet{Boroson_2010a} classified 102 of these observations
as hard states and used them to demonstrate the orbital variability of
absorption in the hard state, but did not have enough data taken
during softer states. At the same time, higher sensitivity snapshots
at different $\phi_{\mathrm{orb}}$ with Chandra
\citep{Miller_2005a,Hanke_2009a,Hanke_2011_PhD,Miskovicova_2014a} and
Suzaku \citep{Nowak_2011a, Miller_2012a, Yamada_2013a} revealed a
complex structure of the wind with highly variable local densities,
but the orbital coverage was significantly worse than the RXTE
measurements.

\begin{figure}
\resizebox{\hsize}{!}{\includegraphics{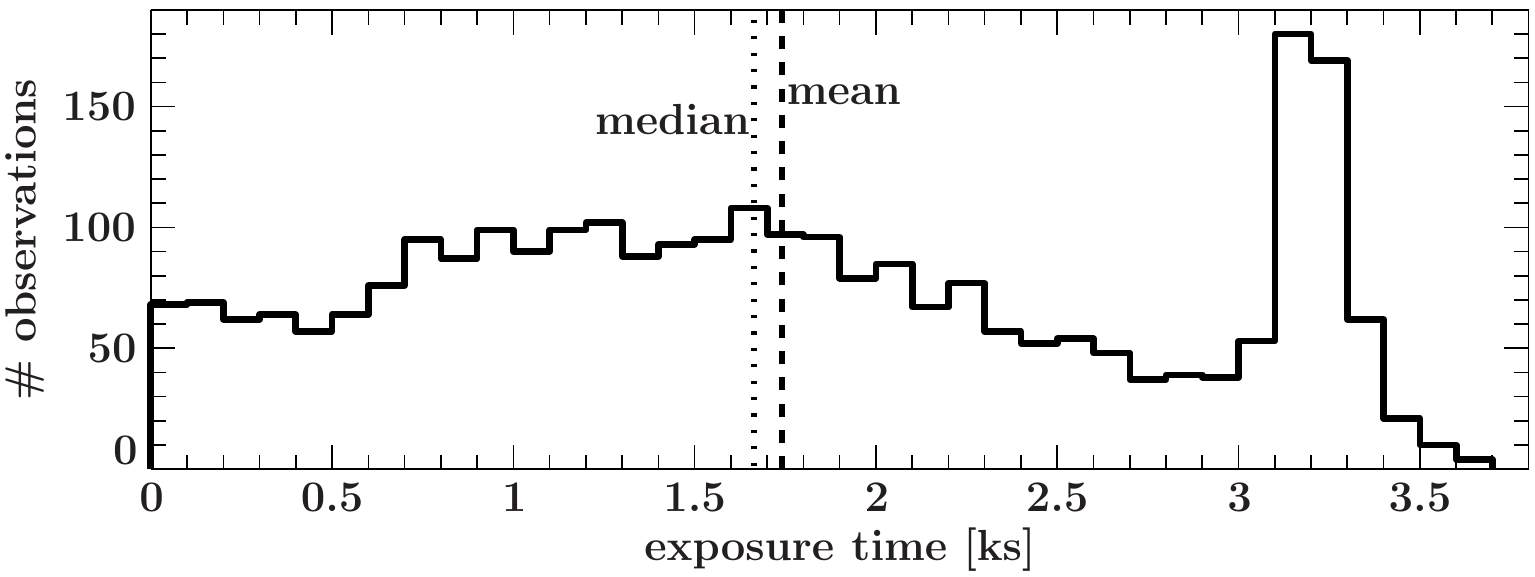}}
\caption{Histogram of the PCA exposure times of the individual
  satellite orbit-wise observations analyzed in this work. The mean
  exposure is 1.74\,ks, the median 1.66\,ks.}\label{fig:exposures}
\end{figure}

\begin{figure}
\resizebox{\hsize}{!}{\includegraphics{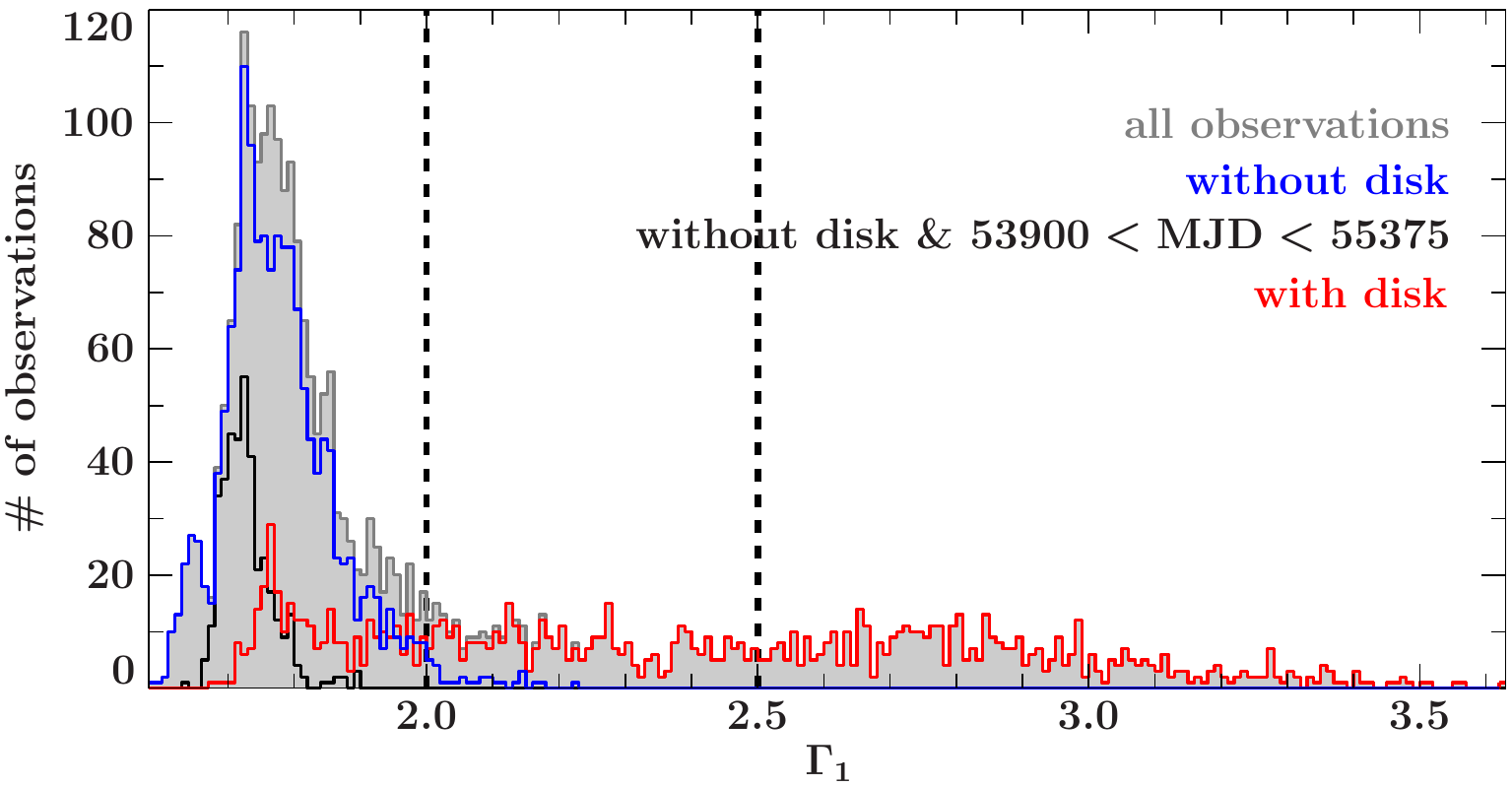}}
\caption{Number of orbitwise PCA observations with a given soft photon
  index $\Gamma_1$ that do (red) or do not (blue) require a
  multitemperature disk component. Additionally highlighted are
  observations that do not require a disk and fall into the long hard
  state of MJD 53900--55375 (black). Dashed vertical lines indicate
  thresholds for hard ($\Gamma_1 \leq 2.0$), intermediate ($2.0 <
  \Gamma_1 \leq 2.5$) and soft ($2. 5 < \Gamma_1$) states according to
  \citet{Grinberg_2013a}.}\label{fig:gamma}
\end{figure}

In this paper we therefore extend the earlier RXTE analyses, using the
full mission-long database of RXTE observations of Cyg~X-1, and using
spectra taken with the exposure of one orbit of the RXTE satellite
around Earth, or around 1.6\,ks (Fig.~\ref{fig:exposures}). We thus
probe average spectral parameters on shorter timescales than earlier
RXTE studies over a variety of spectral shapes
(Fig.~\ref{fig:gamma}). Our total sample consists of 2741 PCU~2 and
HEXTE-spectra \citep{Grinberg_2013a}. The total exposure time of our
observations is 4.77\,Ms which corresponds to almost 10 fully covered
orbital periods of 5.6\,d. The exposure is evenly distributed between
the orbital phases. Advances in PCA calibration
\citep{Shaposhnikov_2012a} and a careful consideration of
energy-dependent systematic uncertainties in different gain epochs of
the PCA-instrument \citep{Hanke_2011_PhD,Grinberg_2013a} strengthen
our analysis. For a detailed discussion of data treatment, see
\citet{Grinberg_2013a}. All analyses in this paper were performed with
ISIS~1.6.2 \citep{Houck_Denicola_2000a,Houck_2002,Noble_Nowak_2008a}.

\begin{figure}
\resizebox{\hsize}{!}{\includegraphics{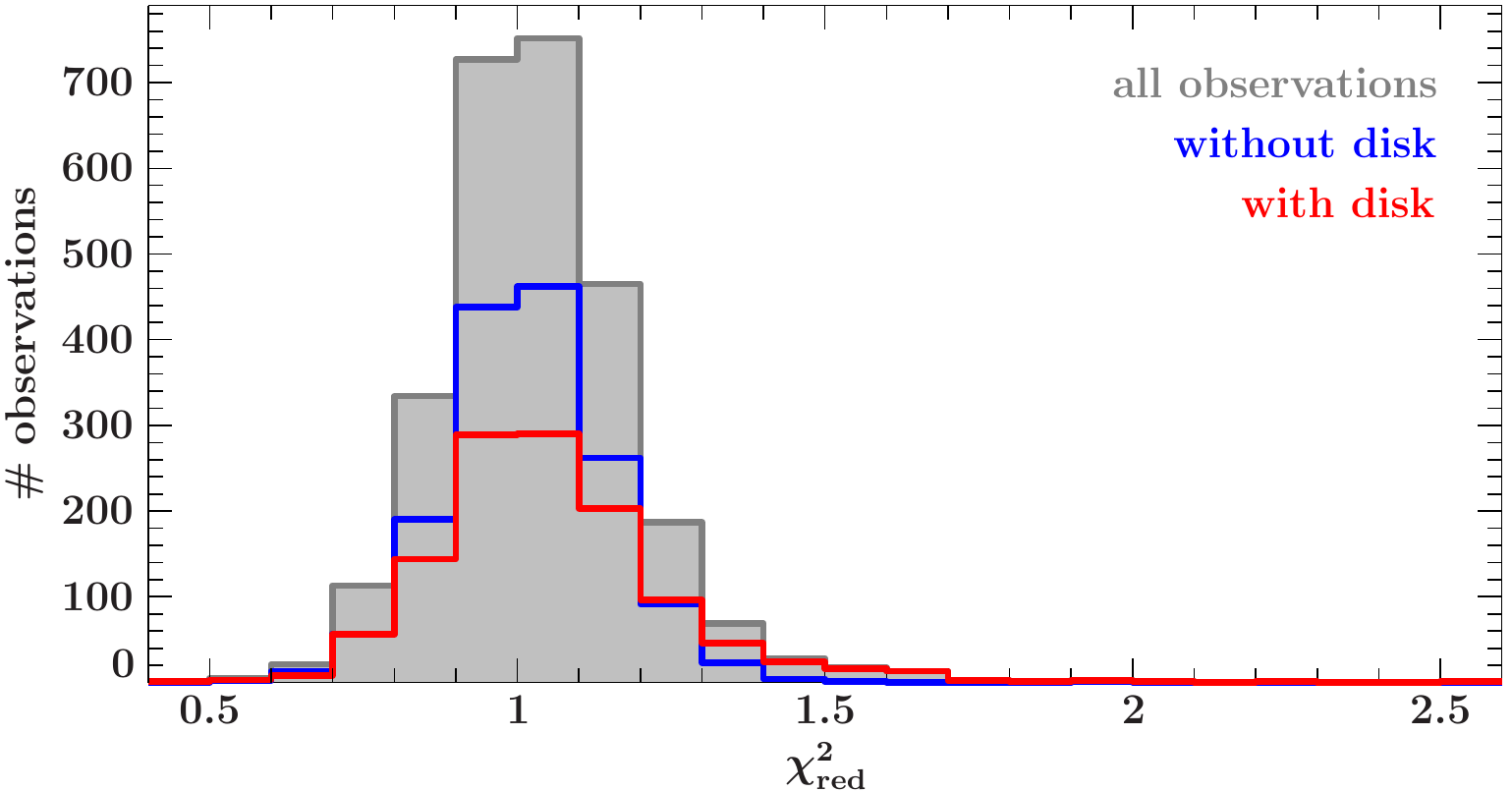}}
\caption{Histogram of $\chi^2_{\mathrm{red}}$ values for observations
  that do (red) and do not (blue) require a multitemperature disk
  component.}\label{fig:chi}
\end{figure}

In performing our analysis we used the same approach as that employed
by us previously \citep{Grinberg_2013a}, i.e., we model the data
empirically with a broken power law with soft photon index $\Gamma_1$,
hard photon index $\Gamma_2$, and a spectral break at $\sim$10\,keV
\citep{Wilms_2006a}. This broken power law is modified by a
high-energy cutoff, by an Fe K$\alpha$-line modelled with a Gaussian
at $\sim$6.4\,keV, and by absorption. A multitemperature disk
\citep[\texttt{diskbb},][]{Mitsuda_1984a,Makishima_1986a} is added to
the model where it improves the $\chi^2$ by more than 5\% except in
seven cases where the X-ray timing behavior and the correlation
between $\Gamma_1$ and $\Gamma_2$ strongly prefer the model without a
disk \citep{Grinberg_2014a}. Figure~\ref{fig:chi} shows that the model
offers a very good description of the spectrum. As in
\citet{Grinberg_2013a}, we use $\Gamma_1$-based state definitions
derived from spectral and timing properties of Cyg~X-1: the source is
in hard state if $\Gamma_1 \leq 2.0$, in the intermediate state if
$2.0 < \Gamma_1 \leq 2.5$ and in the soft state if $2. 5 < \Gamma_1$.
The distribution of observations with $\Gamma_1$ is shown in
Fig.~\ref{fig:gamma}. Out of 1822 hard state observations, 1515 do not
require a disk, but 384 out of 413 intermediate state observations and
all 506 soft state observations require a disk component.

When studying variations of $N_\mathrm{H}$ in an astronomical source,
the measured column towards the source consists of two main
components: absorption in the interstellar medium (ISM) that is
constant on observable timescales and a variable absorption local to
the X-ray source. For Galactic sources, X-ray data do not allow us to
distinguish the two absorption components. In the case of Cyg~X-1, we
therefore take the constant ISM contribution into account by setting a
minimum value for $N_{\mathrm{H}}$ to $N_{\mathrm{H,ISM}} = 4.8\times
10^{21}\,\mathrm{cm}^{-2}$, as derived by \citet{Xiang_2011a} from
studies of the dust scattering halo of the source. We describe
absorption with the \texttt{tbnew} model, an improved
version\footnote{\url{http://pulsar.sternwarte.uni-erlangen.de/wilms/research/tbabs/}}
of \texttt{tbabs} \citep{Wilms_2000a}, using \texttt{wilm} abundances
of \citet{Wilms_2000a} and \texttt{vern} cross sections of
\citet{Verner_1996a}. Note that optical depth at a frequency $\nu$
depends on the total cross-section $\sigma_\nu$ and the hydrogen
column density $N_{\mathrm{H}}$ as $\tau_\nu = \sigma_\nu \cdot
N_{\mathrm{H}}$. The observed spectrum, $I_{\mathrm{obs}}$, and the
source spectrum, $I_{\mathrm{src}}$, are then related by
$I_{\mathrm{obs},\nu} = I_{\mathrm{src},\nu}\cdot e^{-\tau_\nu}$.  For
the individual contribution to $\sigma_\nu$ from different phases of
the absorbing medium (e.g., gas or dust) and the different
constituents (e.g., oxygen) see \citet{Wilms_2000a} or
\citet{Gatuzz_2015a}. Generally in the X-ray range, the optical depth
is dominated by metals \citep{Wilms_2000a} and $\sigma_\nu \propto
E^{-3}$, i.e., the photoelectric absorption acts more strongly on the
low energy photons.

We caution that the exact choice of absorption
model, abundances, and cross-sections can lead to differences in
$N_{\mathrm{H}}$ up to 20\%--30\% between different approaches
\citep{Wilms_2000a}, although they do not affect overall trends
observed. We also caution that the absorption modeling is based on the
absorption of neutral material, while in reality the material will be
moderately photoionized by the strong X-ray source embedded in
it. Especially when using proportional counters such as the PCUs,
approximating the X-ray absorption by ionized material with absorption
in neutral material is an appropriate assumption, since most
absorption is due to K-shell absorption, which is only mildly
dependent on ionization stage. However, our modelling will not pick up
effects from fully ionized material, as it is mainly
transparent to X-rays and RXTE's instruments did not have the
resolution to detect absorption lines.

A second issue relates to the uncertainty of the $N_\mathrm{H}$
measurements. For models such as the ones used here there is a well
known degeneracy between the power law slope, i.e., $\Gamma_1$, and
$N_{\mathrm{H}}$ \citep[e.g.,][]{Suchy_2008a}. By calculating
two-dimensional confidence contours we confirm that this degeneracy is
not responsible for the variation of $N_{\mathrm{H}}$ discussed here
\citep[for example contours see][]{Hanke_2011_PhD}.  Because of the
limited energy range of the PCA, in addition to the correlation with
the power law continuum there is also a degeneracy between parameters
of the accretion disk and $N_\mathrm{H}$. The effect of a stronger
absorption can be counterweighted by a stronger disk component and
vice versa. $N_\mathrm{H}$-values from best fit models with and
without a disk therefore have to be considered separately because of
the different systematic effects and we do so throughout this
paper. Unless discussed otherwise, we give all uncertainties at the
90\% level.

\begin{figure}
\resizebox{\hsize}{!}{\includegraphics{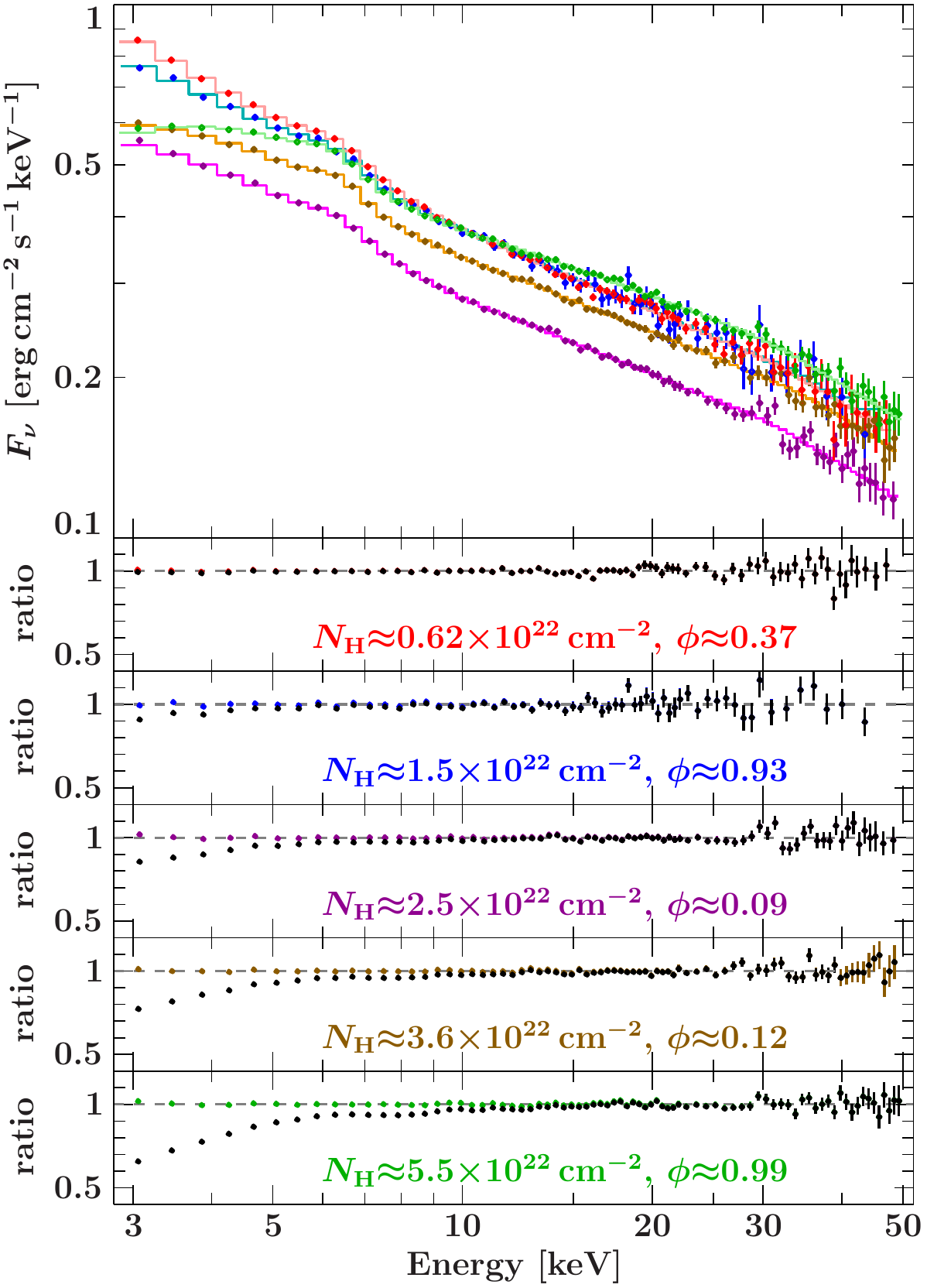}}
\caption{Example spectra with $1.72 < \Gamma_1 < 1.73$, but different
  absorption columns. PCA data and best fit models to PCA and HEXTE
  data are shown. The ratio plots show residuals of the best fit in
  color and residuals of an evaluation of the best fit model with
  $N_{\mathrm{H}}=N_{\mathrm{H,ISM}}$ (i.e., no absorption local to
  the system) in black. All observations are from the hard state of
  2006--2010 (Sect.~\ref{sect:analysis_hard}).}\label{fig:nhs}
\end{figure}

\section{Orbital variability of absorption}\label{sect:orbvar}

\subsection{Hard state}\label{sect:analysis_hard}

\begin{figure}
\includegraphics[width=\columnwidth]{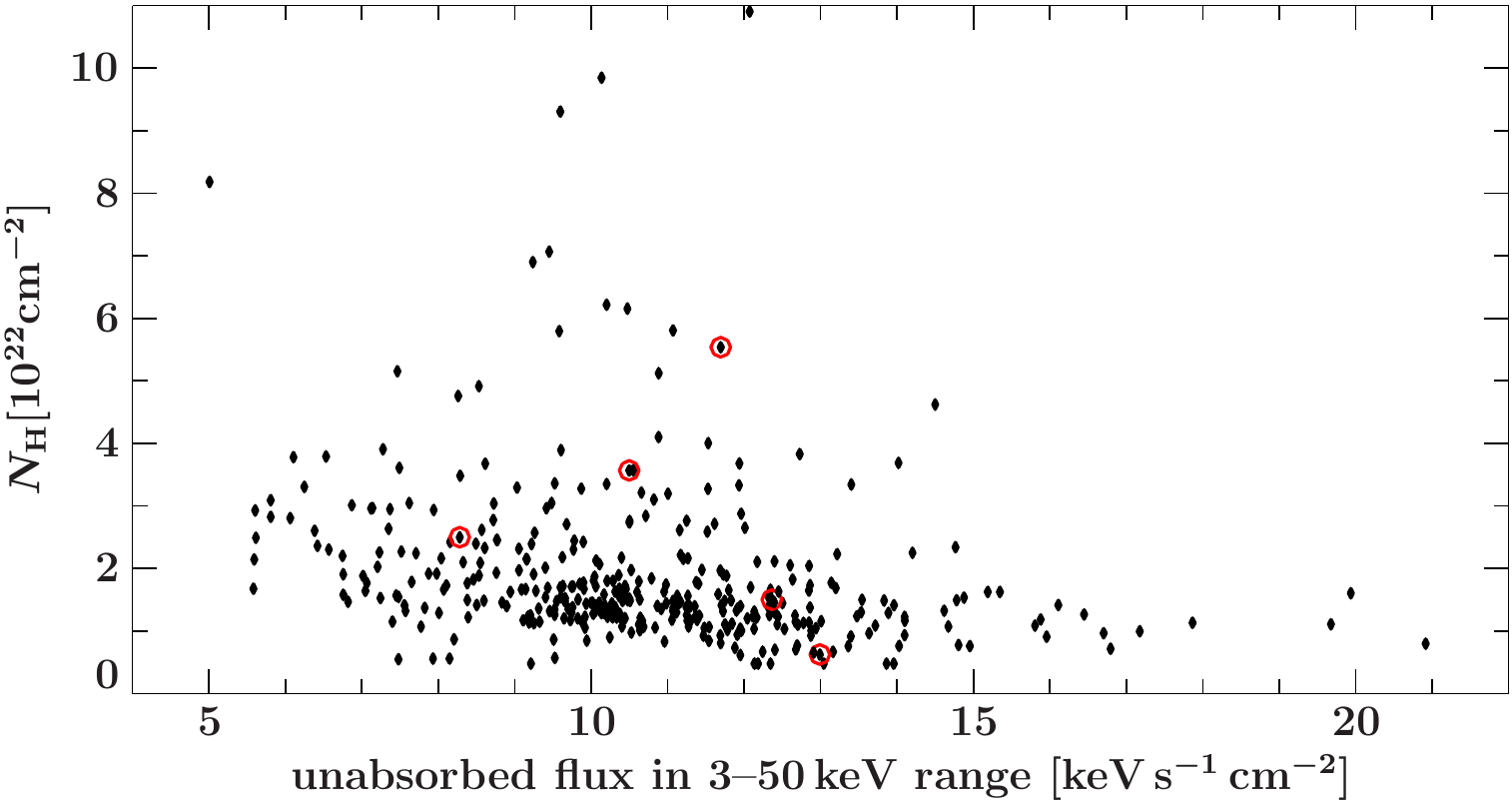}
\caption{Absorption column vs. unabsorbed energy flux during the hard 
  state of 2006--2010. Red circles mark observations shown in 
  Fig.~\ref{fig:nhs}.}\label{fig:flux}
\end{figure}

We now turn to studying the variability of $N_\mathrm{H}$ in the RXTE
data. As an example, Fig.~\ref{fig:nhs} shows spectra of different
observations which have the same typical hard state continuum shape
($\Gamma_1=1.72\ldots 1.73$) but different absorption columns. The
residuals shown in the figure for various orbital phases illustrate
the strong effect caused by the orbital variation compared to the
influence of the interstellar medium alone. Absorption affects the
whole spectrum up to energies of 10\,keV. 

\begin{figure*}
\includegraphics[width=0.32\textwidth]{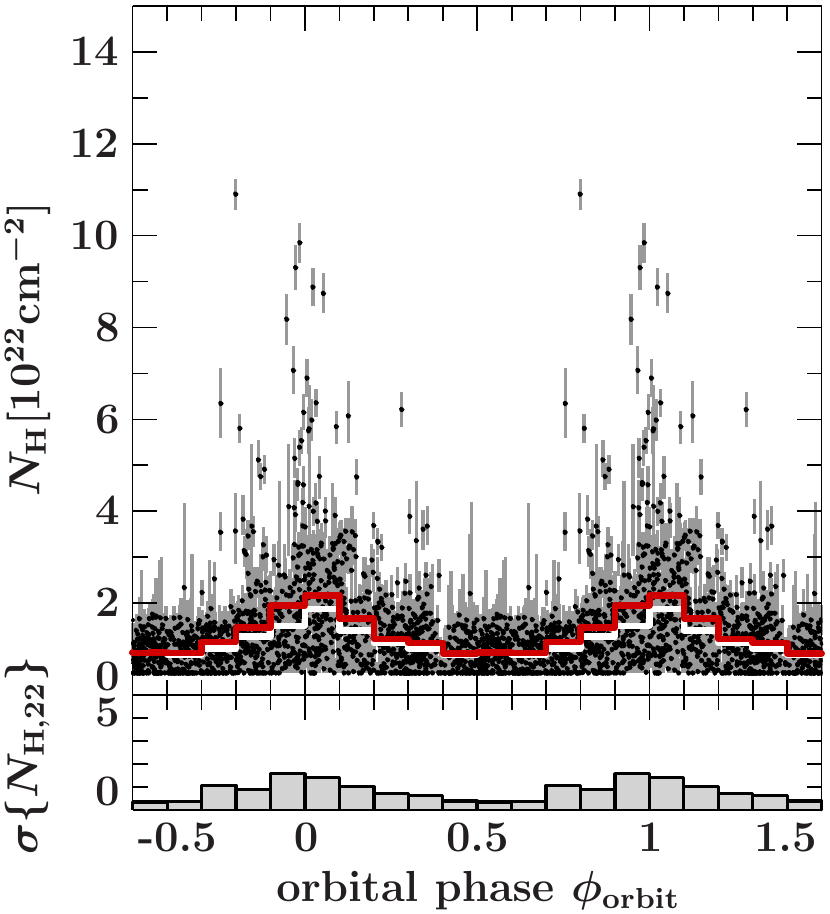}\hfill 
\includegraphics[width=0.32\textwidth]{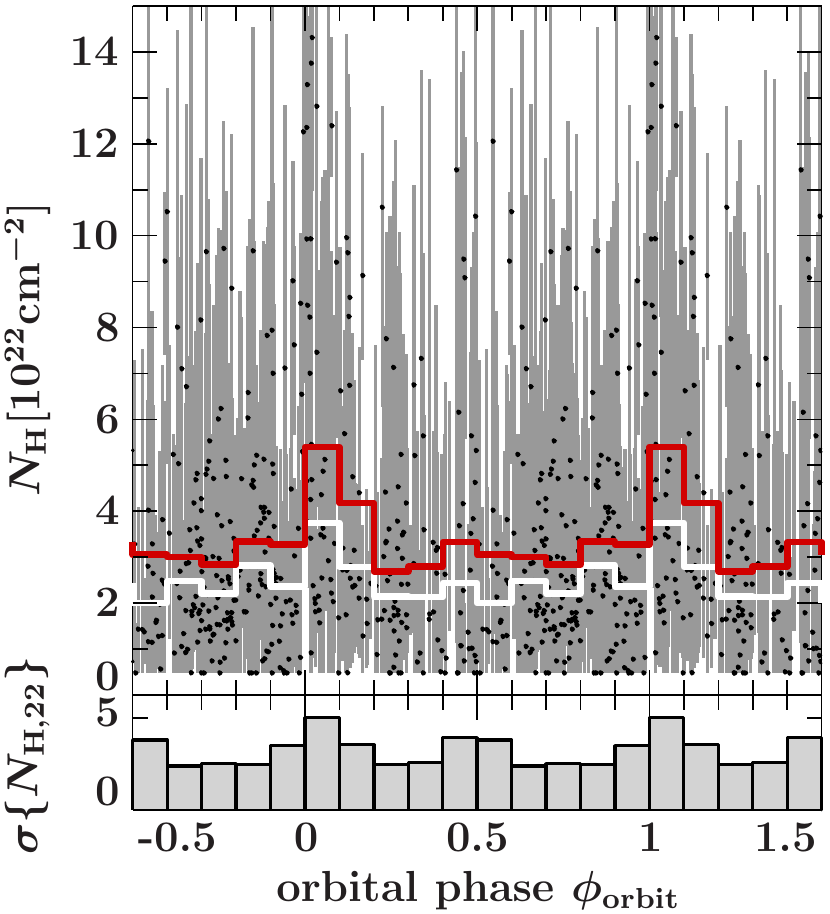}\hfill 
\includegraphics[width=0.32\textwidth]{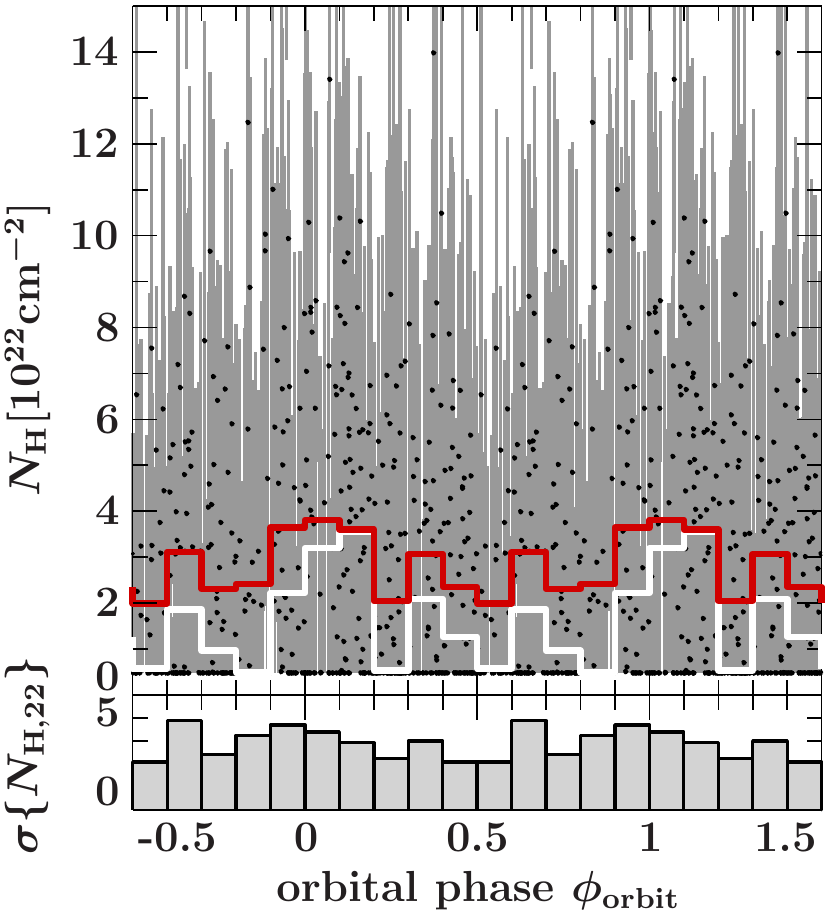}
\caption{Variation of $N_\mathrm{H}$ with orbital phase for
  observations where the source was in the hard (left), intermediate
  (middle), and soft (right) state. The upper panels show the orbit
  dependence of $N_\mathrm{H}$ measurements for individual satellite
  orbit-wise spectra (black datapoints with gray uncertainties), as
  well as the average (red) and median (white) values for orbital phase
  bins with a width of $\Delta \phi_{\mathrm{orb}} = 0.1$. The lower
  panels show the standard deviation $\sigma \{N_{\mathrm{H,22}}\}
  \coloneqq \sigma \{N_\mathrm{H}/(10^{22}\,\mathrm{cm}^{-2})\}$ of
  the $N_\mathrm{H}$ measurements. The panels specifically show 1515
  hard state ($\Gamma_1 \leq 2.0$) observations that do not require a
  disk component, 384 intermediate state ($2.0 < \Gamma_1 \leq 2.5$)
  observations that require a disk, and 506 soft state ($2.5 <
  \Gamma_1$) observations (all soft state observations require a disk
  component).}\label{fig:nhorbit}
\end{figure*}

When looking at the whole state-resolved sample, we determine the
average and median values of $N_\mathrm{H}$ and its standard
deviation, $\sigma \{N_{\mathrm{H,22}}\} \coloneqq \sigma
\{N_\mathrm{H}/(10^{22}\,\mathrm{cm}^{-2})\}$, for phase bins of width
$\Delta \phi_{\mathrm{orb}} = 0.1$ starting at $\phi_{\mathrm{orb}} =
0$. In the hard state, the average column values and the standard
deviation, i.e., the variability of the wind between individual
observations, increase around superior conjunction
(Fig.~\ref{fig:nhorbit}, left).

Individual observations, especially for $\phi_{\mathrm{orb}} \sim 0$,
show $N_\mathrm{H}$ values much larger than the average for a given
phase (Fig.~\ref{fig:nhorbit}, left, see also Fig.~\ref{fig:nhs} for
example spectra). We associate these observations with the prolonged
deep absorption events (often dubbed ``dips'') known from the
literature (Sect.~\ref{sect:intro}) and carefully check our data to
find out whether these observations have systematic differences to
other observations in the same spectral state. We find no such
differences; in particular, these ``dips'' do not have larger
uncertainties in $N_\mathrm{H}$ and do not show clear trends in their
exposure time, $\chi^2_{\mathrm{red}}$, unabsorbed 3--50\,keV
flux\footnote{The X-ray flux of Cyg~X-1 can vary by factor of 4--5 at
  the same spectral shape \citep[][and references
  therein]{Wilms_2006a}.} (see also Fig.~\ref{fig:flux}), or the
correlation between $\Gamma_1$ and $\Gamma_2$ known from
\citet{Wilms_2006a}. The dips cluster around superior conjunction, but
individual strong absorption measurements occur as early as at
$\phi_{\mathrm{orb}} \approx 0.8$ and as late as at
$\phi_{\mathrm{orb}} \approx 0.3$ (Fig.~\ref{fig:nhorbit}, left).

The RXTE spectra do not allow a meaningful fit of partial covering
models and so we cannot constrain possible partial covering either in
physical space or in time (i.e., a clump or a group of clumps,
possibly of different optical depth, passing through the line of sight
for a part of the total exposure time of one spectrum). Thus, the
$N_\mathrm{H}$-values derived for individual observations are lower
limits on the sum of the absorption columns of individual clumps with
the ISM absorption over a given observation's exposure time (i.e., on
average 1.6--1.7\,ks, Fig.~\ref{fig:exposures}).

\begin{figure*}
  \hfill\includegraphics[height=0.2\textheight]{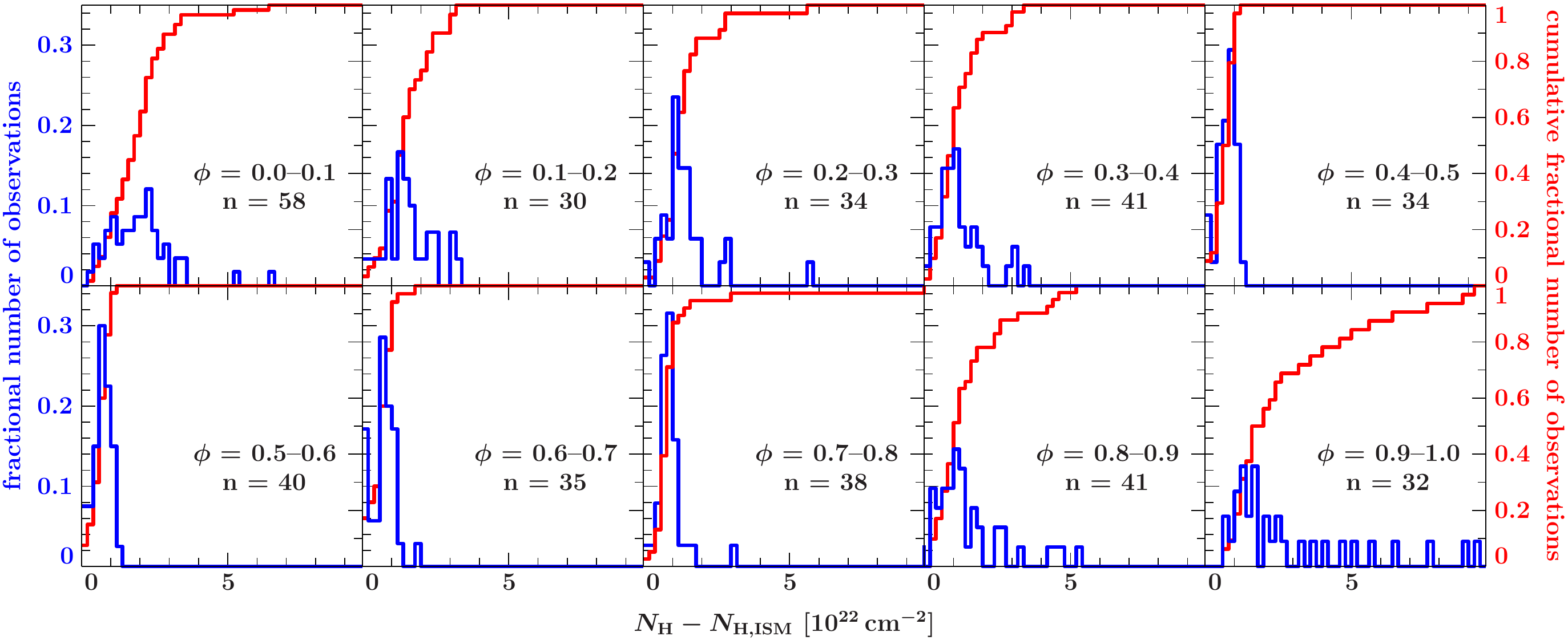}\hfill
\includegraphics[height=0.2\textheight]{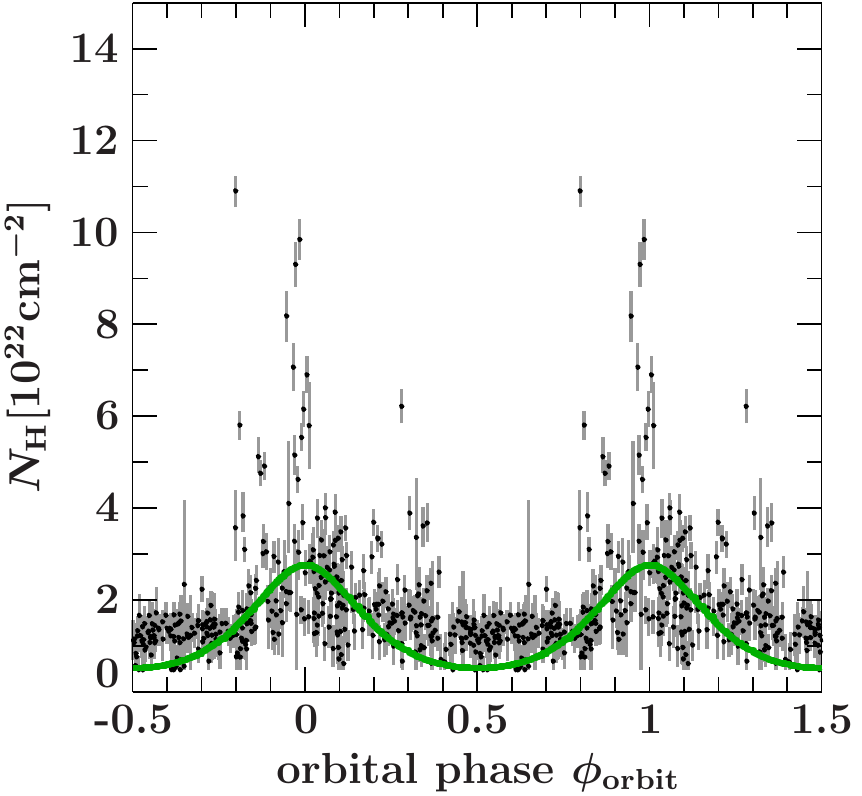}\hfill 
\caption{Observations from the hard state of 2006--2010
  \citep[MJD~53900--55375,][]{Grinberg_2013a} that do not require a
  disk. \textsl{Left}: normalized histograms (blue) and normalized
  cumulative histograms (red) for $N_\mathrm{H}$ above the ISM value
  for individual orbital phase bins; $n$ denotes the total number of
  observations in a given orbital phase bin. Note that the phase bin
  $\phi = 0.7\texttt{--}0.8$ contains one measurement in excess of
  $10^{23}\,\mathrm{cm}^{-2}$. \textsl{Right}: Orbital dependence of
  $N_\mathrm{H}$. The best fit focussed wind model
  introduced in Sect.~\ref{sect:cak} is shown in
  green.}\label{fig:longhard}
\end{figure*}

Within the hard state, Cyg X-1 can display a range of spectral and
timing behaviors
\citep{Pottschmidt_2003b,Wilms_2006a,Grinberg_2014a}. At the same
time, we expect the behavior of the absorption to depend on the
broadband spectral shape of the X-ray source because of possible
ionization of the wind. For a detailed analysis of absorption, we
therefore need a hard state with stable spectral and timing
characteristics, such as the long hard state of 2006--2010
\citep[MJD~53900--55375,][]{Grinberg_2013a,Grinberg_2014a}. We show
results for this long hard state period in Fig.~\ref{fig:longhard}; it
includes 383 observations that do not require a disk
(Fig.~\ref{fig:gamma}) and that have a total exposure time of 660\,ks
or $\sim$1.4 orbital periods of the system. The variability of
absorption can be clearly seen in the (cumulative) probability
distributions of $N_\mathrm{H}$ values at different orbital phases
(Fig.~\ref{fig:longhard}, left). The values are low and the
distribution narrow at $\phi_{\mathrm{orb}} \sim 0.5$. At
$\phi_{\mathrm{orb}} \sim 0$, the distribution is broad with a higher
average (average values and their standard deviations are shown in
Fig.~\ref{fig:theory}, right). We will use this set for comparisons
with a toy model for the focussed wind and a clumpy wind model in
Sect.~\ref{sect:cak} and~\ref{sect:clumpy}, respectively.

\subsection{Intermediate state and soft state}

We now address observations of Cyg~X-1 in the intermediate and the
soft state, i.e., when the X-ray spectrum is much softer than in the
hard state. Most data in the intermediate state require a disk
component that leads to larger uncertainties in the $N_\mathrm{H}$
determination and particularly to a systematically increased average
value independent of orbital phase as well as a larger $\sigma
\{N_{\mathrm{H,22}}\}$. An increase of average $N_\mathrm{H}$ at
$\phi_{\mathrm{orb}} \approx 0$ is, however, still visible
(Fig.~\ref{fig:nhorbit}, middle).

Our data do not show orbital variability of absorption in the soft
state (Fig.~\ref{fig:nhorbit}, right). The average values are
comparable with the intermediate state, but in several phase bins we
find median values of $4.8\times 10^{21}\,\mathrm{cm}^{-2}$, i.e., the
data are consistent with no neutral absorption intrinsic to the
source. Because of the increased disk contribution, observations in
the soft state are even more affected by systematics than those in the
intermediate state, especially given that the disk in Cyg X-1 has a
temperature $\lesssim$0.4\,keV ($\approx 4.6 \times 10^6$\,K) even in
the soft state \citep[][and our fits]{Wilms_2006a}, i.e., it peaks
below the lower limit of our data so that the disk parameters show
relatively large uncertainties.

\section{Discussion}\label{sect:disc}

\subsection{Hard state}
\subsubsection{Discussion and comparison to earlier results}
Our results show that despite PCA's
energy coverage only above $\sim$3\,keV and comparatively low spectral
resolution when compared to CCD or even grating instruments, PCA data
can be used to assess trends in the behavior of the absorption. The
changes in $N_\mathrm{H}$ strongly influence the shape of the observed
spectra (Fig.~\ref{fig:nhs}). The overall orbital variability of
$N_\mathrm{H}$ values found here is consistent with earlier results
\citep{Feng_2002a,Ibragimov_2005a,Wilms_2006a,Boroson_2010a}.
To our knowledge, we are the first to quantify the variance of the
equivalent hydrogen column, $\sigma \{N_{\mathrm{H,22}}\}$
(Fig.~\ref{fig:nhorbit}, lower panels).

The largest $N_\mathrm{H}$ values seen during the hard state
(Figs.~\ref{fig:nhs} and \ref{fig:longhard}) are on the order of
$10^{23}\,\mathrm{cm}^{-2}$. \citet[][using
  \textsl{Tenma}]{Kitamoto_1984a}, \citet[][using
  \textsl{ASCA}]{Balucinska-Church_1997a}, and \citet[][using
  \textsl{Chandra}]{Hanke_2008a} reported individual deep absorption
events with $N_\mathrm{H} \gtrsim 10^{23}\,\mathrm{cm}^{-2}$. Periods
of increased absorption that lasted longer than a few ks have been
observed with various instruments, especially close to superior
conjunction. For example, \citet{Feng_2002a} discuss a long dip of at
least $\sim$3\,ks that shows complex substructure. Prolonged periods
of increased absorption are also clearly seen in the long
uninterrupted observations possible with \textsl{Chandra}
\citep{Hanke_2008a}. Given these previous measurements and the size of
our sample, the detection of several strong absorption events as
discussed in Sect.~\ref{sect:analysis_hard} is plausible.

\citet{Balucinska-Church_2000a} have used mainly RXTE-All Sky Monitor
\citep[ASM;][]{Levine_1996a} data to show orbital variability in
ASM-based hardness ratios and in dip (defined by a threshold ASM
hardness ratio) occurrence, both peaking at superior conjunction
consistent with the behavior of the absorption column density that we
observe. They observe a peak in the dip occurence at
$\phi_{\mathrm{orb}} = 0.95$, but our sample is not large enough to
track this possible assymetry of the $N_\mathrm{H}$ distribution
around $\phi_{\mathrm{orb}} = 0$. \citet{Poutanen_2008a} further show
that the orbital modulation of ASM countrates in the hard state
depends on the superorbital phase: the number of ASM-defined dips
shows a superorbital periodicity when assuming a $\sim$150\,d
superorbital period. The smaller number of our measurements when
compared to ASM data combined with the apparent stochastic nature and
large variance of the absorption variations do not allow such an
analysis here.

Despite these shortcomings, the exceptional coverage of the pointed
observation allows us to compare the observational results to
theoretical models. In particular we quantitatively compare a simple
focussed CAK wind to our data from the stable hard state period of
2006--2010 ($\sim$250 orbits of the binary system) in
Sect.~\ref{sect:cak}, and, for the first time for a high mass black
hole binary, qualitatively match the observed data from the same
stable hard state period with predictions from a clumpy wind model for
O-star winds in Sect.~\ref{sect:clumpy}.

\subsubsection{Focussed wind model in the hard state}\label{sect:cak}

We first compare the orbital variation of $N_\mathrm{H}$ from MJD
53900--55370 with a toy model for the focussed wind as introduced by
\citet{Gies_1986b}. This model is based on the previous work of
\citet{Friend_1982a} and depends on the angle $\theta$ between the
wind direction and the donor-black hole axis.  We emphasize that this
is only a toy model that does not take into account details of both
the physics of a real wind and the observations used. The model is,
however, useful to assess general trends and to point out the
directions in which better models have to be developed.

\begin{figure}
\resizebox{\hsize}{!}{\includegraphics{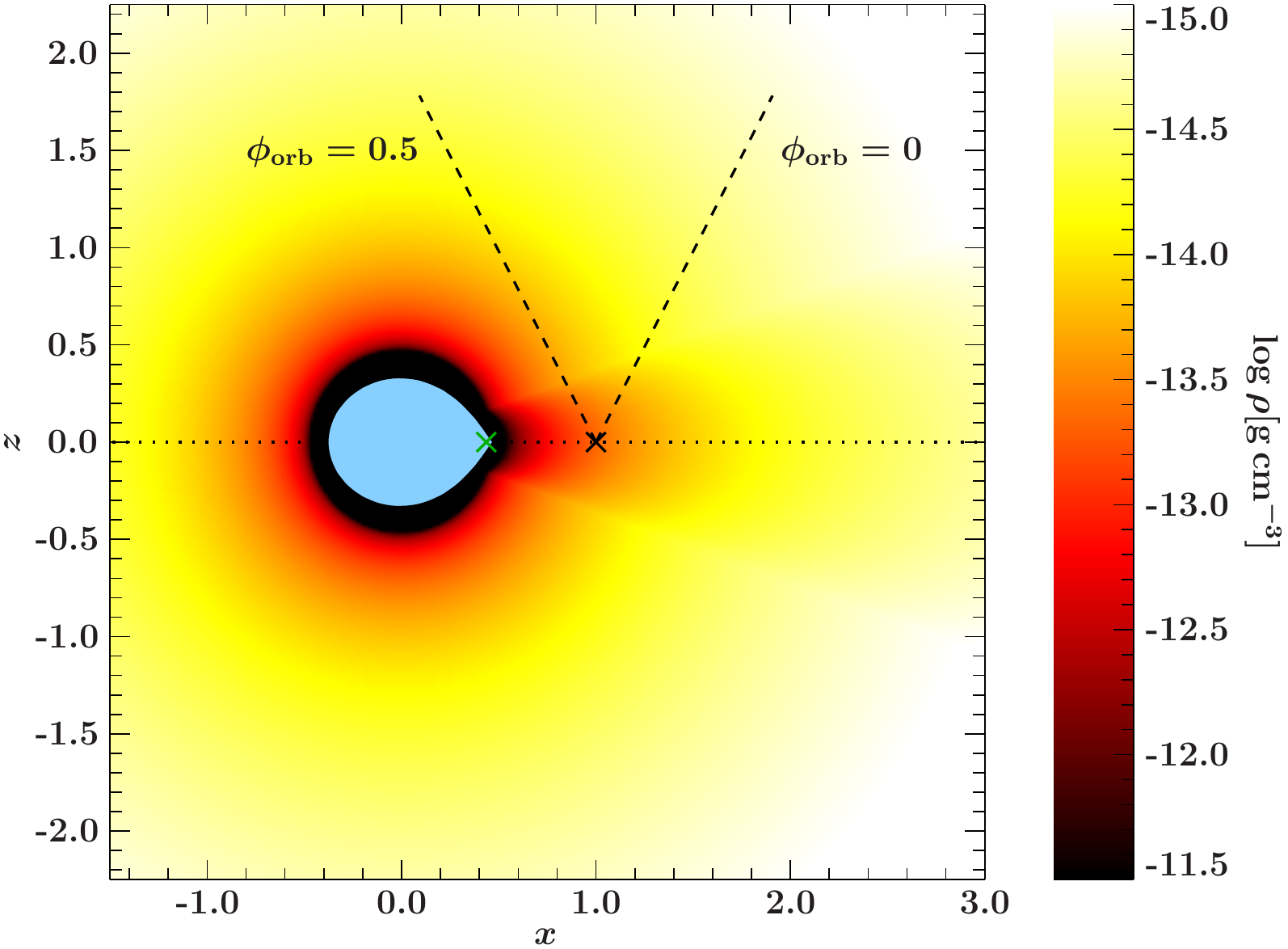}}
\caption{Stellar wind density calculated following the focussed wind
  model of \citet{Gies_1986a}. The $x$- and $z$- axes are shown in units of
  orbital separation. The position of the black hole is marked with a
  black cross, the position of the center of mass with a green
  cross. Lines of sight towards the black hole at orbital phases 0 and
  0.5 are indicated by dashed lines, the orbital plane by a dotted
  line. Note that there are no discotinuities in the wind; if such
  appear above, they are unfortunate tricks of
  vision.}\label{fig:density}
\end{figure}

For $\theta \geqq 20^\circ$, the wind is the radiatively driven
(CAK-)wind of a single star \citep{Castor_1975a}. For $\theta <
20^\circ$, the wind varies smoothly with $\theta$ with terminal
velocity and density peaking at $\theta = 0$. In particular, we employ
the model with the fill-out-factor of 0.98 with the correspondent
values for terminal velocity and wind density listed in Table~1 of
\citet{Gies_1986a} and use masses, inclination, and orbital period as
given in Sect.~\ref{sect:intro}.  Figure ~\ref{fig:density} shows the
resulting distribution of wind density.

We obtain the total (neutral and ionized) column density by
integrating along the line of sight towards the black hole at a given
orbital phase: these values are on the order of
3--$5\times10^{22}\mathrm{cm}^{-2}$, i.e., much higher than the
measured average values and more reminiscent of dips
(cf. Fig.~\ref{fig:nhorbit}, left) But this total column density is
not the $N_{\mathrm H}$ measured in our observations: the fully
ionized part of the wind is transparent to X-rays and thus only some
weakly to moderately ionized part of the wind will contribute to the
measured absorption. We can hence assume that the absorption in the
binary is restricted to the region of the wind with $\log \xi < \log
\xi_{\mathrm {max}}$ with the ionization parameter $\xi$ defined after
\citet{Tarter_1969a} as $\xi = L/(nr^2)$ with $n$ being the absorbing
particle number density, $r$ the distance from the ionizing source,
and $L$ the luminosity above the hydrogen Lyman edge \citep[typical
for Cyg~X-1: $L = 10^{37}\,\mathrm{erg}\,\mathrm{s}^{-1}$,
][]{Miskovicova_2014a}.  We then consider only the contribution of
those parts of the wind that satisfy this condition.

We take into account the ISM column density $N_{\mathrm{H,ISM}}$ and
fit the parameter $\log \xi_{\mathrm {max}}$ of this model to the
data. We obtain $\log \xi_{\mathrm {max}} \approx 2.7$, however with
$\chi^2_{\mathrm{red}} = 9.1$ for 382\,degrees of freedom. The
  phase variation of $N_\mathrm{H}$ that corresponds to our best fit
  value of $\log \xi_{\mathrm {max}}$ is shown in
  Fig.~\ref{fig:longhard}, right panel, as a green curve. The failure
to describe the data well is expected since the toy model does not
include clumping that we expect to majorly contribute to the spread of
the $N_{\mathrm{H}}$ measurements at a given orbital phase. To conduct
a simple test for the influence of clumping on our results, we remove
the 13 measurements with $N_\mathrm{H} > 5 \times
10^{22}\,\mathrm{cm}^{-2}$ and obtain a bestfit $\chi^2_{\mathrm{red}}
= 4.5$ for 369\,degrees of freedom with a decreased bestfit value
($\log \xi_{\mathrm {max}} \approx 2.67$ as compared to previously
$\log \xi_{\mathrm {max}} \approx 2.71$).

The toy model also neglects additional smaller effects such as a
potential photoionization wake that would introduce asymmetries
\citep{Blondin_1994a}. Uncertainties in determination of the ISM
absorption, $N_{\mathrm{H,ISM}}$, towards the source may lead to
further problems \citep{Xiang_2011a}. However, it is interesting to
note that when we account for the fact that our measurements include
the clumping, our result is in rough agreement with results ($\log \xi
\lesssim 2.6$) found by \citet{Miskovicova_2014a}, who infer
$N_\mathrm{H}$ values ranging from $6.8 \times
10^{21}\,\mathrm{cm}^{-2}$ to $1.3 \times 10^{22}\,\mathrm{cm}^{-2}$
from modelling the Ne edge in the non-dip parts of hard state
\textsl{Chandra} observations at 5 different orbital phases.

\subsubsection{Clumpy wind model for the hard state}\label{sect:clumpy}

\begin{figure*}
\hfill\includegraphics[height=0.2\textheight]{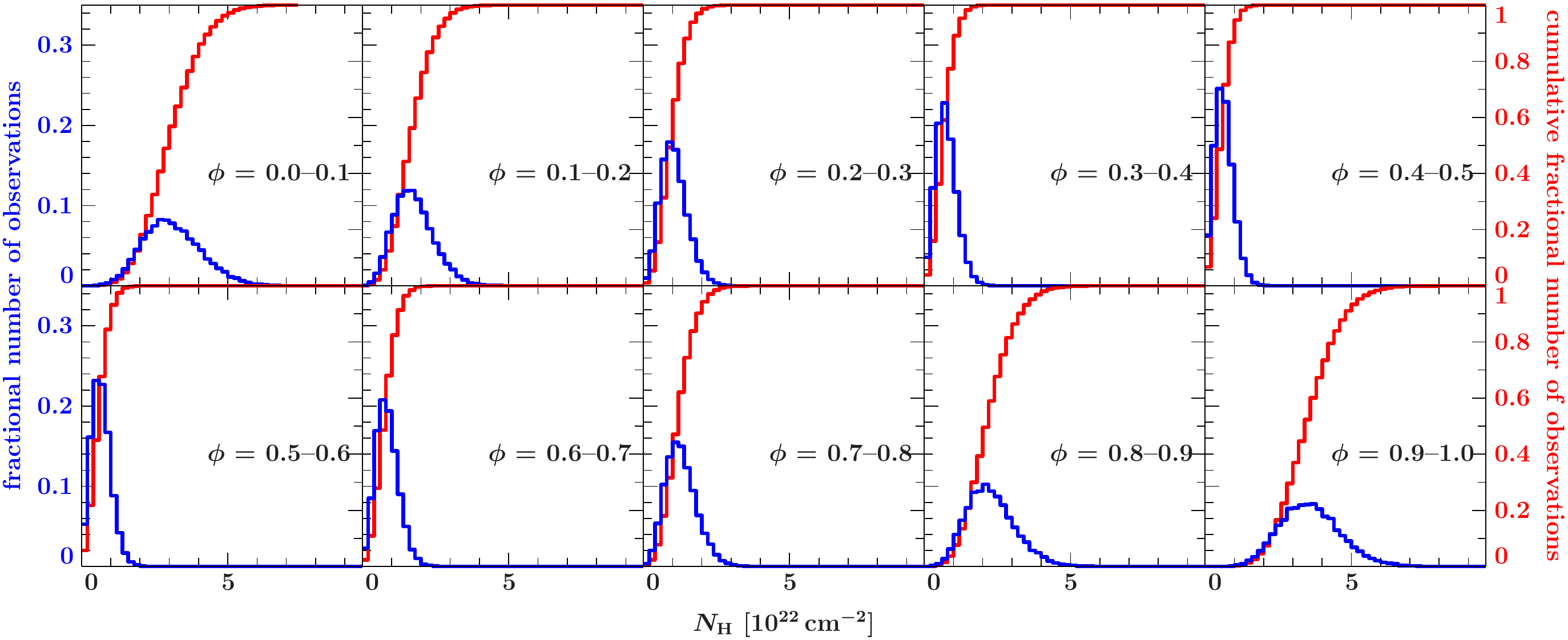}\hfill
\includegraphics[height=0.2\textheight]{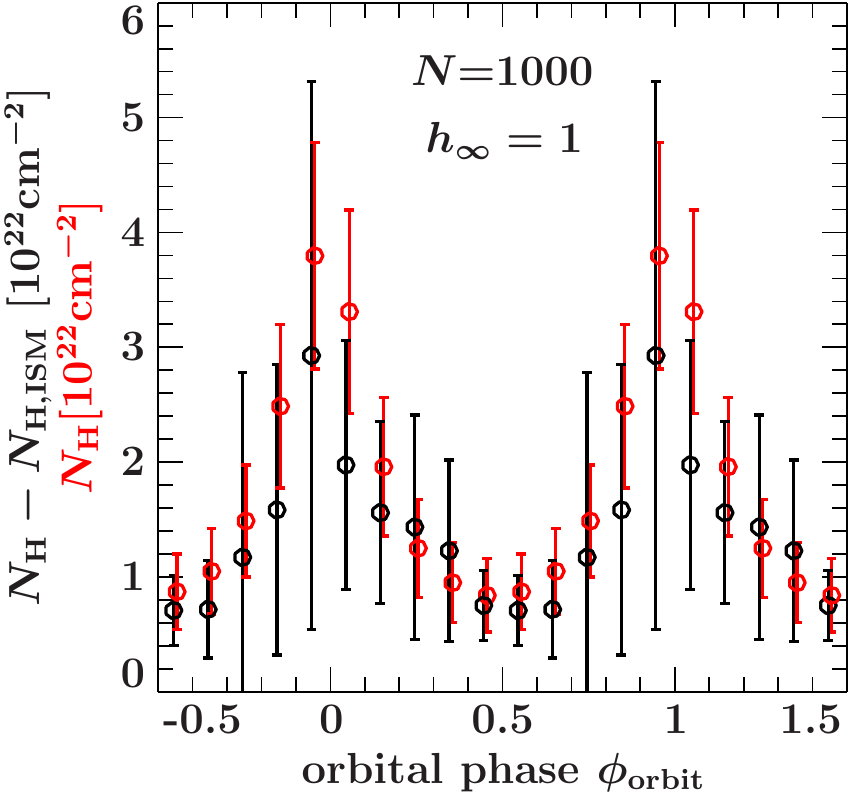}\hfill
\caption{Predictions of $N_\mathrm{H}$ variability assuming a clumpy
  wind model with the total number of clumps per flow time $N=1000$
  and the terminal porosity length $h_\infty = R_*$ (see
  Sect.~\ref{sect:clumpy} for explanation of
  parameters). \textsl{Left}: model histograms for the distribution of
  equivalent absorption column density $N_{\mathrm{H}}$.
  \textsl{Right}: average values (circles) and standard deviations
  (`error bars' on the average values) in a given orbital phase for
  the theoretical model (red) and values measured in the hard state of
  MJD 53900-55375 (black). Average measured values are shown as
  `$\mbox{average} - N_{\mathrm{H,ISM}}$' to account for the ISM
  absorption.}\label{fig:theory}
\end{figure*}

We next consider whether the observed variability in absorbing column
could be due to wind structures, and how the characteristics of those
structures compare to what is known about radiatively driven winds in
other contexts, such as clumps \citep{Abbott_1981a, Hillier_1998a,
  Fullerton_2006a, Puls_2006a} or discrete absorption components
\citep[DACs;][]{Prinja_1986a}. To this end, we use a numerical code to
simulate the column density between the observer and a point-like
X-ray source, as the source orbits around a companion whose wind is
composed of discrete clumps. For simplicity, the code does not account
for the focussed wind properties of the system; moreover, the clumps
are assumed to be spherical, rather than the pancake morphologies
considered by other authors
\citep[e.g.,][]{Feldmeier_2003a,Oskinova_2004a}. At a given orbital
phase, the column-density evaluation is performed by casting a ray
from the source to the observer, determining which ray segments lie
inside clumps, and summing up the contributions from these segments
(as the product of the segment length and the clump density).

The distribution of clumps throughout the wind follows a similar
approach to that described in Appendix A of
\citet{Sundqvist_2012b}. Clumps are relased at the stellar surface
$r=R_{*}$ in a random direction and with an initial radial size
$l_{*}$, and then advect outward with the wind according to a
canonical `$\beta$' velocity law $\varv(r) = \varv_{\infty} (1 -
R_{*}/r)^{\beta}$. As a clump advects its radial size grows as $l(r) =
(r/R_{*}) l_{*}$, but its mass remains fixed at $m = \dot{M}/\dot{N}$
where $\dot{M}$ is the overall wind mass loss rate, and $\dot{N}$ is
the clump release rate. The latter is most conveniently parameterized
in terms of the number of clumps per wind flow time $N = \dot{N}
t_{\rm flow}$, with $t_{\rm flow} \equiv R_{*} / \varv_{\infty}$. To
relate these parameters to the porosity length $h$ \citep[essentially,
the mean free path between clumps; see][]{Owocki_2006a}, we note that
$h(r) = h_{\infty} \varv(r)/\varv_{\infty}$, where the terminal
porisity length is
\begin{equation}
h_{\infty} \equiv 3 \frac{R_{*}}{l_{*}^{2} N}.
\end{equation}

To apply the code to Cyg X-1, we assume $\beta = 1.6$, $R_{*} =
18\,R_{\odot}$, $\varv_{\infty} = 2400\,{\rm km\,s^{-1}}$, and
$\dot{M} = 2 \times 10^{-6}\,{\rm M_{\odot}\,yr^{-1}}$, and adopt the
orbital parameters given in Section 1. We ran a grid of simulations
with $N = 100, 316, 1000, 3162$ and 10000, and $h_{\infty}/R_{*} =
0.1, 0.316, 1, 3.162$, and 10 to explore the level of variability
expected on the time scales we probe here. For each grid point we
calculated 1000 model orbits, with a phase resolution $\Delta
\phi_{\mathrm{orb}} = 0.001$. Since typical RXTE observations of Cyg
X-1 had an exposure of about 0.005 orbits, and since each observation
is not subdivided into time slices, it is necessary to average the
model to a timescale corresponding to the observations. Thus, we
rebinned the models by a factor of five by taking a linear average of
the model column density. A linear average over a phase bin does not
strictly correctly account for the non-linear addition of partial
covering in time, but is adequate given the crude nature of our
model. Likewise, rebinning by a factor of five is not an exact match
to the actual observations, but it is an adequate approximation.

The statistical properties of each phase bin of 0.005 orbits do not
change rapidly as a function of orbital phase, so these statistical
properties were further averaged into phase bins of 0.1 orbits,
corresponding to the grouping used to evaluate the observations.  We
show this statistical analysis of the column for $N =1000$ and
$h_\infty = R_*$ in Fig.~\ref{fig:theory}; the mean and standard
deviation of this model are compared to those of the data in the right
panel. We have chosen a model with a similar level of variability as
the observations near phase 0.5, where our line of sight to the
compact object mainly probes the part of the wind that is likely least
perturbed by the peculiarities of the focussed part of the wind. For
other model realizations see Figs.~\ref{fig:theory_all} and
\ref{fig:std_all}.

The choice of porosity length $h_\infty$ has a strong effect in
determining the level of variability in the model
(Fig.~\ref{fig:theory_all}). The choice of $N$ has a weaker effect,
with larger values of $N$ (and thus small values of $l_*$) showing
somewhat reduced variability (Fig.~\ref{fig:std_all}). This is a
result of single observations averaging over the transit of one or
more clumps across the line of sight, thus washing out the variability
expected on short timescales. We have explored the effect of different
choices of averaging timescale (or equivalently observation duration)
in our simulations, and as expected, we find that increasing the
averaging timescale tends to reduce the variability (see also
Sect.~\ref{sect:analysis_hard} for a discussion of exposure times on
the measured $N_{\mathrm{H}}$). We also find that there is a maximum
useful time resolution above which increasing the sampling rate does
not increase the level of variability observed. This maximum rate
corresponds roughly to the time scale for a typical clump to transit
our line of sight to the compact object which is on the order of
$l_*/\varv_\infty$ (when the wind velocity dominates over orbital
velocity as is the case here). One obvious conclusion of these results
is that the clump size scale could be constrained by obtaining spectra
of high statistical quality (e.g., with \textsl{Suzaku} or
\textsl{Chandra} satellites) and exploring the level of variability on
shorter time scales.

The models with $h_\infty = R_*$ have the best agreement with the data
near $\phi_{\mathrm{orb}} = 0.5$, with the exception that for $N =
10000$ the model with $h_\infty = 3.162 R_*$ has the best agreement
(Fig.~\ref{fig:std_all}). Based on previous theoretical and
observational investigations of single O stars, it is reasonable that
$h_\infty$ should be comparable to or smaller than $R_*$
\citep{Dessart_2003a, Owocki_2006a, Naze_2013a}.  This is also
consistent with the results of \citet{Leutenegger_2013a}, who found
that the spectrum of the single O supergiant $\zeta$ Pup could be fit
with models having porosity lengths $h_\infty \leq R_*$.  To test for
the influence of terminal velocity on our results, we re-ran the
simulations with $\varv_{\infty} =
1600\,\mathrm{km}\,\mathrm{s}^{-1}$. Although the resulting mean
values for $N_{\mathrm{H}}$ are higher (due to the increased wind
density), our results are otherwise similar to the original
$\varv_{\infty} = 2400\,\mathrm{km}\,\mathrm{s}^{-1}$ grid, and thus
our conclusions do not change.

Comparing Fig.~\ref{fig:longhard} and~\ref{fig:theory}, the model
shows a Gaussian distribution of absorbing columns, while the data
show a non-Gaussian tail near phase 0. It seems likely that this
non-Gaussian tail reflects some unusual wind structure related to the
focussed wind. However, since the impact parameter of our line of sight
with respect to the optical companion is smallest near this orbital
phase, the tail might also be a signature of a wind with
pancake-shaped clumps, which would be effectively more porous when
viewed edge on \citep{Feldmeier_2003a,Oskinova_2012a}. A more thorough
exploration of these effects may give insight into their importance
but it is beyond the scope of this paper.

\subsection{Intermediate state and soft state}
Due to the lack of suitable observations in the previously analyzed
samples, the variability of absorption in intermediate and soft states
has not been previously analyzed using RXTE spectra, but only using
ASM data. \citet{Boroson_2010a} analyzed orbital variability of
absorption in five soft state periods. \citeauthor{Boroson_2010a} see
no variability in their soft states~2 and~5 that would be mostly
classified as intermediate states using the more rigorous state
definitions of \citet{Grinberg_2013a}. Their soft state~3 shows
orbital variability. \citet{Boroson_2010a} also point out that the
intrinsic variability of the X-ray source in the softer periods
hampers the search for periodicities in ASM data.

Based on the pointed RXTE data alone, we see no orbital variability
in the soft state, while there are signs for an increase of the main
absorption toward $\phi_{\mathrm{orb}} \approx 0$ in the intermediate
state. These results qualitatively agree with the behavior seen in
analyses of \textsl{Chandra} high resolution spectra: as the source
softens, dipping becomes less pronounced \citep{Miskovicova_2014a}.
However, dips have still been observed in the soft state with
\textsl{Suzaku} \citep[][but note the strong changes in hardness ratio
  in their data that may be indicative of spectral evolution during
  the observation]{Yamada_2013a}. \citet{Hanke_2011_PhD} shows
lightcurves and softness ratios of all \textsl{Chandra} observations
up to and including ObsID~13219 taken in 2011~February: some weak and
short dipping can be seen in the soft state data. 

The interpretation of these results in the context of the previous
\textsl{Suzaku} and \textsl{Chandra} results is that the soft X-rays
emitted from the vicinity of the black hole in the soft state are
strong enough to fully ionize the wind and even some or all of the
optically thick clumps. The wind thus becomes (mainly) transparent to
X-rays and no orbital variation of absorption column can be seen.

\section{Summary and conclusions}\label{sect:sum}

We have analyzed almost 5\,Ms of RXTE observations of Cyg X-1 taken
over a span of 16 years in different spectral states. We find clear
orbital variability of absorption in the hard state of the X-ray
source with an overall increase and additional deep absorption events
(``dips'' with $N_{\mathrm H} \sim 10^{23} \mathrm{cm}^{-2}$) around
superior conjunction, in agreement with previous work with different
instruments. Because of the presence of an additional multitemperature
disk component in the spectra, the absorption column cannot be well
constrained in the intermediate and soft states. However, there are
signs for increased $N_\mathrm{H}$ towards $\phi_{\mathrm{orb}} = 0$
in the intermediate state.

We compared the observed orbital variability of absorption with two
wind models. A toy model for a focussed line-driven wind without
clumps \citep{Gies_1986b} does not describe the data well and we
attribute the differences to the absence of clumping in the model that
would lead to the observed strong variations of $N_\mathrm{H}$ at the
same orbital phase. Qualitatively matching the observed $N_\mathrm{H}$
distributions at a given orbital phase to a toy clumpy wind model
\citep{Owocki_2006a,Sundqvist_2012b} without a focussed wind and with
spherical clumps results in the best match for models with porosity
length, $h_\infty$, on the order of the stellar radius, $R_*$, at
orbital phases $\phi_{\mathrm{orb}} \approx 0.5$, in agreement with
results from (effectively) single O stars
\citep{Leutenegger_2013a}. Our models exclude both much lower and much
higher porosities. The discrepancies between the toy clumpy
  wind model and data at $\phi_{\mathrm{orb}} \approx 0$ could be due
to either a focussed wind component or a non-spherical shape of the
clumps or a combination of both. We note here that the
  existence of a focussed wind component in Cyg X-1 is supported by
  multiple lines of evidence such as optical measurements
  \citep{Gies_1986b,Sowers_1998a}.

While we caution against an over-interpretation of details of the
presented simple models they firstly demonstate that RXTE data of Cyg
X-1 can indeed be used for an analysis of the long-term orbital
variability of absorption and secondly clearly demand better
models. In particular, a thorough exploration of the winds will
require a model that includes both, the focussed and the clumpy
structure of the wind, and possibly also different clump shapes and/or
sizes.

\begin{acknowledgements} 
  Support for this work was provided by NASA through the Smithsonian
  Astrophysical Observatory (SAO) contract SV3-73016 to MIT for
  Support of the Chandra X-Ray Center (CXC) and Science Instruments;
  CXC is operated by SAO for and on behalf of NASA under contract
  NAS8-03060. It was partially completed by LLNL under Contract
  DE-AC52-07NA27344, and is supported by NASA grants to LLNL and
  NASA/GSFC. We thank the Bundesministerium f\"ur Wirtschaft und
  Technologie for funding through Deutsches Zentrum f\"ur Luft- und
  Raumfahrt grant 50 OR 1113. MAN acknowledges support from NASA Grant
  NNX12AE37G. RHDT acknowledges support from NASA award
  NNX12AC72G. This research has made use of NASA's Astrophysics Data
  System Bibliographic Services. We thank John E.  Davis for the
  development of the \texttt{slxfig} module used to prepare all
  figures in this work and Fritz-Walter Schwarm, Thomas Dauser, and
  Ingo Kreykenbohm for their work on the Remeis computing
  cluster. This research has made use of ISIS functions
  (\texttt{isisscripts}) provided by ECAP/Remeis observatory and
  MIT\footnote{\url{http://www.sternwarte.uni-erlangen.de/isis/}}.
  Without the hard work by Evan Smith and Divya Pereira to schedule
  the observations of \mbox{Cyg~X-1} so uniformly for more than a
  decade, this whole series of papers would not have been possible.
\end{acknowledgements}

%BIBLIOGRAPHY
%\bibliographystyle{jwaabib} 
\bibliographystyle{aa} 
\bibliography{mnemonic,aa_abbrv,references}

\begin{thebibliography}{87}
\expandafter\ifx\csname natexlab\endcsname\relax\def\natexlab#1{#1}\fi

\bibitem[{{Abbott} {et~al.}(1981){Abbott}, {Bieging}, \&
  {Churchwell}}]{Abbott_1981a}
{Abbott}, D.~C., {Bieging}, J.~H., \& {Churchwell}, E. 1981, \apj, 250, 645

\bibitem[{{Axelsson} {et~al.}(2006){Axelsson}, {Borgonovo}, \&
  {Larsson}}]{Axelsson_2006a}
{Axelsson}, M., {Borgonovo}, L., \& {Larsson}, S. 2006, \aap, 452, 975

\bibitem[{{Ba{\l}uci{\'n}ska-Church} {et~al.}(2000){Ba{\l}uci{\'n}ska-Church},
  {Church}, {Charles}, {Nagase}, {LaSala}, \&
  {Barnard}}]{Balucinska-Church_2000a}
{Ba{\l}uci{\'n}ska-Church}, M., {Church}, M.~J., {Charles}, P.~A., {et~al.}
  2000, \mnras, 311, 861

\bibitem[{{Ba{\l}uci{\'n}ska-Church} {et~al.}(1997){Ba{\l}uci{\'n}ska-Church},
  {Takahashi}, {Ueda}, {Church}, {Dotani}, {Mitsuda}, \&
  {Inoue}}]{Balucinska-Church_1997a}
{Ba{\l}uci{\'n}ska-Church}, M., {Takahashi}, T., {Ueda}, Y., {et~al.} 1997,
  \apjl, 480, L115

\bibitem[{Belloni(2010)}]{Belloni_2010a}
Belloni, T.~M. 2010, in The Jet Paradigm: From Microquasars to Quasars, Lecture
  Notes in Physics, Berlin Springer Verlag, ed. {T.~Belloni}, Vol. 794, 53

\bibitem[{{Benlloch} {et~al.}(2004){Benlloch}, {Pottschmidt}, {Wilms}, {Nowak},
  {Gleissner}, \& {Pooley}}]{Benlloch_2004a}
{Benlloch}, S., {Pottschmidt}, K., {Wilms}, J., {et~al.} 2004, in American
  Institute of Physics Conference Series, Vol. 714, X-ray Timing 2003: Rossi
  and Beyond, ed. P.~{Kaaret}, F.~K. {Lamb}, \& J.~H. {Swank}, 61--64

\bibitem[{{Blondin}(1994)}]{Blondin_1994a}
{Blondin}, J.~M. 1994, \apj, 435, 756

\bibitem[{{B{\"o}ck} {et~al.}(2011){B{\"o}ck}, {Grinberg}, {Pottschmidt},
  {Hanke}, {Nowak}, {Markoff}, {Uttley}, {Rodriguez}, {Pooley}, {Suchy},
  {Rothschild}, \& {Wilms}}]{Boeck_2011a}
{B{\"o}ck}, M., {Grinberg}, V., {Pottschmidt}, K., {et~al.} 2011, \aap, 533,
  A8+

\bibitem[{{Boroson} \& {Vrtilek}(2010)}]{Boroson_2010a}
{Boroson}, B. \& {Vrtilek}, S.~D. 2010, \apj, 710, 197

\bibitem[{{Brocksopp} {et~al.}(1999){Brocksopp}, {Fender}, {Larionov}, {Lyuty},
  {Tarasov}, {Pooley}, {Paciesas}, \& {Roche}}]{Brocksopp_1999b}
{Brocksopp}, C., {Fender}, R.~P., {Larionov}, V., {et~al.} 1999, \mnras, 309,
  1063

\bibitem[{{Castor} {et~al.}(1975){Castor}, {Abbott}, \& {Klein}}]{Castor_1975a}
{Castor}, J.~I., {Abbott}, D.~C., \& {Klein}, R.~I. 1975, \apj, 195, 157

\bibitem[{{Davis} \& {Hartmann}(1983)}]{Davis_1983a}
{Davis}, R. \& {Hartmann}, L. 1983, \apj, 270, 671

\bibitem[{{Dessart} \& {Owocki}(2003)}]{Dessart_2003a}
{Dessart}, L. \& {Owocki}, S.~P. 2003, \aap, 406, L1

\bibitem[{{Feldmeier} {et~al.}(2003){Feldmeier}, {Oskinova}, \&
  {Hamann}}]{Feldmeier_2003a}
{Feldmeier}, A., {Oskinova}, L., \& {Hamann}, W.-R. 2003, \aap, 403, 217

\bibitem[{{Feldmeier} {et~al.}(1997){Feldmeier}, {Puls}, \&
  {Pauldrach}}]{Feldmeier_1997a}
{Feldmeier}, A., {Puls}, J., \& {Pauldrach}, A.~W.~A. 1997, \aap, 322, 878

\bibitem[{{Feng} \& {Cui}(2002)}]{Feng_2002a}
{Feng}, Y.~X. \& {Cui}, W. 2002, \apj, 564, 953

\bibitem[{{Friend} \& {Castor}(1982)}]{Friend_1982a}
{Friend}, D.~B. \& {Castor}, J.~I. 1982, \apj, 261, 293

\bibitem[{{Fullerton} {et~al.}(2006){Fullerton}, {Massa}, \&
  {Prinja}}]{Fullerton_2006a}
{Fullerton}, A.~W., {Massa}, D.~L., \& {Prinja}, R.~K. 2006, \apj, 637, 1025

\bibitem[{{Gatuzz} {et~al.}(2015){Gatuzz}, {Garc{\'{\i}}a}, {Kallman},
  {Mendoza}, \& {Gorczyca}}]{Gatuzz_2015a}
{Gatuzz}, E., {Garc{\'{\i}}a}, J., {Kallman}, T.~R., {Mendoza}, C., \&
  {Gorczyca}, T.~W. 2015, \apj, 800, 29

\bibitem[{{Gies} \& {Bolton}(1986{\natexlab{a}})}]{Gies_1986a}
{Gies}, D.~R. \& {Bolton}, C.~T. 1986{\natexlab{a}}, \apj, 304, 371

\bibitem[{{Gies} \& {Bolton}(1986{\natexlab{b}})}]{Gies_1986b}
{Gies}, D.~R. \& {Bolton}, C.~T. 1986{\natexlab{b}}, \apj, 304, 389

\bibitem[{{Gies} {et~al.}(2008){Gies}, {Bolton}, {Blake}, {Caballero-Nieves},
  {Crenshaw}, {Hadrava}, {Herrero}, {Hillwig}, {Howell}, {Huang}, {Kaper},
  {Koubsk{\'y}}, \& {McSwain}}]{Gies_2008a}
{Gies}, D.~R., {Bolton}, C.~T., {Blake}, R.~M., {et~al.} 2008, \apj, 678, 1237

\bibitem[{{Gies} {et~al.}(2003){Gies}, {Bolton}, {Thomson}, {Huang}, {McSwain},
  {Riddle}, {Wang}, {Wiita}, {Wingert}, {Cs{\'a}k}, \& {Kiss}}]{Gies_2003a}
{Gies}, D.~R., {Bolton}, C.~T., {Thomson}, J.~R., {et~al.} 2003, \apj, 583, 424

\bibitem[{{Gleissner} {et~al.}(2004{\natexlab{a}}){Gleissner}, {Wilms},
  {Pooley}, {Nowak}, {Pottschmidt}, {Markoff}, {Heinz}, {Klein-Wolt}, {Fender},
  \& {Staubert}}]{Gleissner_2004b}
{Gleissner}, T., {Wilms}, J., {Pooley}, G.~G., {et~al.} 2004{\natexlab{a}},
  \aap, 425, 1061

\bibitem[{{Gleissner} {et~al.}(2004{\natexlab{b}}){Gleissner}, {Wilms},
  {Pottschmidt}, {Uttley}, {Nowak}, \& {Staubert}}]{Gleissner_2004a}
{Gleissner}, T., {Wilms}, J., {Pottschmidt}, K., {et~al.} 2004{\natexlab{b}},
  \aap, 414, 1091

\bibitem[{{Grinberg} {et~al.}(2013){Grinberg}, {Hell}, {Pottschmidt},
  {B{\"o}ck}, {Nowak}, {Rodriguez}, {Bodaghee}, {Cadolle Bel}, {Case}, {Hanke},
  {K{\"u}hnel}, {Markoff}, {Pooley}, {Rothschild}, {Tomsick}, {Wilson-Hodge},
  \& {Wilms}}]{Grinberg_2013a}
{Grinberg}, V., {Hell}, N., {Pottschmidt}, K., {et~al.} 2013, \aap, 554, A88

\bibitem[{{Grinberg} {et~al.}(2014){Grinberg}, {Pottschmidt}, {B{\"o}ck},
  {Schmid}, {Nowak}, {Uttley}, {Tomsick}, {Rodriguez}, {Hell}, {Markowitz},
  {Bodaghee}, {Cadolle Bel}, {Rothschild}, \& {Wilms}}]{Grinberg_2014a}
{Grinberg}, V., {Pottschmidt}, K., {B{\"o}ck}, M., {et~al.} 2014, \aap, 565, A1

\bibitem[{{Hanke}(2011)}]{Hanke_2011_PhD}
{Hanke}, M. 2011, Dissertation, Universit{\"a}t Erlangen-N{\"u}rnberg

\bibitem[{{Hanke} {et~al.}(2009){Hanke}, {Wilms}, {Nowak}, {Pottschmidt},
  {Schulz}, \& {Lee}}]{Hanke_2009a}
{Hanke}, M., {Wilms}, J., {Nowak}, M.~A., {et~al.} 2009, \apj, 690, 330

\bibitem[{{Hanke} {et~al.}(2008){Hanke}, {Wilms}, {Nowak}, {Schulz},
  {Pottschmidt}, {Lee}, \& {Boeck}}]{Hanke_2008a}
{Hanke}, M., {Wilms}, J., {Nowak}, M.~A., {et~al.} 2008, in Microquasars and
  Beyond

\bibitem[{{Hell} {et~al.}(2013){Hell}, {Mi{\v s}kovi{\v c}ov{\'a}}, {Brown},
  {Wilms}, {Clementson}, {Hanke}, {Beiersdorfer}, {Liedahl}, {Pottschmidt},
  {Porter}, {Kilbourne}, {Kelley}, {Nowak}, \& {Schulz}}]{Hell_2013a}
{Hell}, N., {Mi{\v s}kovi{\v c}ov{\'a}}, I., {Brown}, G.~V., {et~al.} 2013,
  Physica Scripta Volume T, 156, 014008

\bibitem[{{Herrero} {et~al.}(1995){Herrero}, {Kudritzki}, {Gabler}, {Vilchez},
  \& {Gabler}}]{Herrero_1995a}
{Herrero}, A., {Kudritzki}, R.~P., {Gabler}, R., {Vilchez}, J.~M., \& {Gabler},
  A. 1995, \aap, 297, 556

\bibitem[{{Hillier} \& {Miller}(1998)}]{Hillier_1998a}
{Hillier}, D.~J. \& {Miller}, D.~L. 1998, \apj, 496, 407

\bibitem[{{Houck}(2002)}]{Houck_2002}
{Houck}, J.~C. 2002, in High Resolution X-ray Spectroscopy with XMM-Newton and
  Chandra, ed. {G.~Branduardi-Raymont}, published electronically

\bibitem[{{Houck} \& {Denicola}(2000)}]{Houck_Denicola_2000a}
{Houck}, J.~C. \& {Denicola}, L.~A. 2000, in Astronomical Data Analysis
  Software and Systems IX, ed. N.~Manset, C.~Veillet, \& D.~Crabtree, ASP
  Conf.\ Ser.~216, 591

\bibitem[{{Ibragimov} {et~al.}(2005){Ibragimov}, {Poutanen}, {Gilfanov},
  {Zdziarski}, \& {Shrader}}]{Ibragimov_2005a}
{Ibragimov}, A., {Poutanen}, J., {Gilfanov}, M., {Zdziarski}, A.~A., \&
  {Shrader}, C.~R. 2005, \mnras, 362, 1435

\bibitem[{{Ibragimov} {et~al.}(2007){Ibragimov}, {Zdziarski}, \&
  {Poutanen}}]{Ibragimov_2007a}
{Ibragimov}, A., {Zdziarski}, A.~A., \& {Poutanen}, J. 2007, \mnras, 381, 723

\bibitem[{Jahoda {et~al.}(2006)Jahoda, Markwardt, Radeva, Rots, Stark, Swank,
  Strohmayer, \& Zhang}]{Jahoda_2006a}
Jahoda, K., Markwardt, C.~B., Radeva, Y., {et~al.} 2006, ApJS, 163, 401

\bibitem[{{Kitamoto} {et~al.}(1984){Kitamoto}, {Miyamoto}, {Tanaka}, {Ohashi},
  {Kondo}, {Tawara}, \& {Nakagawa}}]{Kitamoto_1984a}
{Kitamoto}, S., {Miyamoto}, S., {Tanaka}, Y., {et~al.} 1984, \pasj, 36, 731

\bibitem[{{Leutenegger} {et~al.}(2013){Leutenegger}, {Cohen}, {Sundqvist}, \&
  {Owocki}}]{Leutenegger_2013a}
{Leutenegger}, M.~A., {Cohen}, D.~H., {Sundqvist}, J.~O., \& {Owocki}, S.~P.
  2013, \apj, 770, 80

\bibitem[{Levine {et~al.}(1996)Levine, Bradt, Cui, Jernigan, Morgan, Remillard,
  Shirey, \& Smith}]{Levine_1996a}
Levine, A.~M., Bradt, H., Cui, W., {et~al.} 1996, ApJ, 469, L33

\bibitem[{{Li} \& {Clark}(1974)}]{Li_1974a}
{Li}, F.~K. \& {Clark}, G.~W. 1974, \apjl, 191, L27

\bibitem[{{Lucy} \& {Solomon}(1970)}]{Lucy_1970a}
{Lucy}, L.~B. \& {Solomon}, P.~M. 1970, \apj, 159, 879

\bibitem[{{Makishima} {et~al.}(1986){Makishima}, {Maejima}, {Mitsuda}, {Bradt},
  {Remillard}, {Tuohy}, {Hoshi}, \& {Nakagawa}}]{Makishima_1986a}
{Makishima}, K., {Maejima}, Y., {Mitsuda}, K., {et~al.} 1986, \apj, 308, 635

\bibitem[{{Mason} {et~al.}(1974){Mason}, {Hawkins}, {Sanford}, {Murdin}, \&
  {Savage}}]{Mason_1974a}
{Mason}, K.~O., {Hawkins}, F.~J., {Sanford}, P.~W., {Murdin}, P., \& {Savage},
  A. 1974, \apjl, 192, L65

\bibitem[{{Miller} {et~al.}(2012){Miller}, {Pooley}, {Fabian}, {Nowak}, {Reis},
  {Cackett}, {Pottschmidt}, \& {Wilms}}]{Miller_2012a}
{Miller}, J.~M., {Pooley}, G.~G., {Fabian}, A.~C., {et~al.} 2012, \apj, 757, 11

\bibitem[{{Miller} {et~al.}(2005){Miller}, {Wojdowski}, {Schulz}, {Marshall},
  {Fabian}, {Remillard}, {Wijnands}, \& {Lewin}}]{Miller_2005a}
{Miller}, J.~M., {Wojdowski}, P., {Schulz}, N.~S., {et~al.} 2005, \apj, 620,
  398

\bibitem[{{Mitsuda} {et~al.}(1984){Mitsuda}, {Inoue}, {Koyama}, {Makishima},
  {Matsuoka}, {Ogawara}, {Suzuki}, {Tanaka}, {Shibazaki}, \&
  {Hirano}}]{Mitsuda_1984a}
{Mitsuda}, K., {Inoue}, H., {Koyama}, K., {et~al.} 1984, \pasj, 36, 741

\bibitem[{{Mi\v skovi\v cov\'a} {et~al.}(2015){Mi\v skovi\v cov\'a}, {Hell},
  {Hanke}, {Nowak}, {Pottschmidt}, {Schulz}, {Grinberg}, {Duro}, {Brown}, \&
  {Wilms}}]{Miskovicova_2014a}
{Mi\v skovi\v cov\'a}, I., {Hell}, N., {Hanke}, M., {et~al.} 2015, \aap,
  submitted

\bibitem[{{Morton}(1967)}]{Morton_1967a}
{Morton}, D.~C. 1967, \apj, 150, 535

\bibitem[{{Muijres} {et~al.}(2012){Muijres}, {Vink}, {de Koter}, {M{\"u}ller},
  \& {Langer}}]{Muijres_2012a}
{Muijres}, L.~E., {Vink}, J.~S., {de Koter}, A., {M{\"u}ller}, P.~E., \&
  {Langer}, N. 2012, \aap, 537, A37

\bibitem[{{Naz{\'e}} {et~al.}(2013){Naz{\'e}}, {Oskinova}, \&
  {Gosset}}]{Naze_2013a}
{Naz{\'e}}, Y., {Oskinova}, L.~M., \& {Gosset}, E. 2013, \apj, 763, 143

\bibitem[{Noble \& Nowak(2008)}]{Noble_Nowak_2008a}
Noble, M.~S. \& Nowak, M.~A. 2008, PASP, 120, 821

\bibitem[{{Nowak} {et~al.}(2011){Nowak}, {Hanke}, {Trowbridge}, {Markoff},
  {Wilms}, {Pottschmidt}, {Coppi}, {Maitra}, {Davis}, \&
  {Tramper}}]{Nowak_2011a}
{Nowak}, M.~A., {Hanke}, M., {Trowbridge}, S.~N., {et~al.} 2011, \apj, 728, 13

\bibitem[{{Orosz} {et~al.}(2011){Orosz}, {McClintock}, {Aufdenberg},
  {Remillard}, {Reid}, {Narayan}, \& {Gou}}]{Orosz_2011a}
{Orosz}, J.~A., {McClintock}, J.~E., {Aufdenberg}, J.~P., {et~al.} 2011, \apj,
  742, 84

\bibitem[{{Oskinova} {et~al.}(2004){Oskinova}, {Feldmeier}, \&
  {Hamann}}]{Oskinova_2004a}
{Oskinova}, L.~M., {Feldmeier}, A., \& {Hamann}, W.-R. 2004, \aap, 422, 675

\bibitem[{{Oskinova} {et~al.}(2012){Oskinova}, {Feldmeier}, \&
  {Kretschmar}}]{Oskinova_2012a}
{Oskinova}, L.~M., {Feldmeier}, A., \& {Kretschmar}, P. 2012, \mnras, 421, 2820

\bibitem[{{Owocki} {et~al.}(1988){Owocki}, {Castor}, \&
  {Rybicki}}]{Owocki_1988a}
{Owocki}, S.~P., {Castor}, J.~I., \& {Rybicki}, G.~B. 1988, \apj, 335, 914

\bibitem[{{Owocki} \& {Cohen}(2006)}]{Owocki_2006a}
{Owocki}, S.~P. \& {Cohen}, D.~H. 2006, \apj, 648, 565

\bibitem[{{Parsignault} {et~al.}(1976){Parsignault}, {Epstein}, {Grindlay},
  {Schreier}, {Schnopper}, {Gursky}, {Brinkman}, {Heise}, {Schrijver}, \&
  {Tanaka}}]{Parsignault_1976a}
{Parsignault}, D.~R., {Epstein}, A., {Grindlay}, J., {et~al.} 1976, \apss, 42,
  175

\bibitem[{{Pottschmidt} {et~al.}(2003){Pottschmidt}, {Wilms}, {Nowak},
  {Pooley}, {Gleissner}, {Heindl}, {Smith}, {Remillard}, \&
  {Staubert}}]{Pottschmidt_2003b}
{Pottschmidt}, K., {Wilms}, J., {Nowak}, M.~A., {et~al.} 2003, \aap, 407, 1039

\bibitem[{{Poutanen} {et~al.}(2008){Poutanen}, {Zdziarski}, \&
  {Ibragimov}}]{Poutanen_2008a}
{Poutanen}, J., {Zdziarski}, A.~A., \& {Ibragimov}, A. 2008, \mnras, 389, 1427

\bibitem[{{Pravdo} {et~al.}(1980){Pravdo}, {White}, {Becker}, {Kondo}, {Boldt},
  {Holt}, {Serlemitsos}, \& {McCluskey}}]{Pravdo_1980a}
{Pravdo}, S.~H., {White}, N.~E., {Becker}, R.~H., {et~al.} 1980, \apjl, 237,
  L71

\bibitem[{{Priedhorsky} {et~al.}(1983){Priedhorsky}, {Terrell}, \&
  {Holt}}]{Priedhorsky_1983a}
{Priedhorsky}, W.~C., {Terrell}, J., \& {Holt}, S.~S. 1983, \apj, 270, 233

\bibitem[{{Prinja} \& {Howarth}(1986)}]{Prinja_1986a}
{Prinja}, R.~K. \& {Howarth}, I.~D. 1986, \apjs, 61, 357

\bibitem[{{Puls} {et~al.}(2006){Puls}, {Markova}, {Scuderi}, {Stanghellini},
  {Taranova}, {Burnley}, \& {Howarth}}]{Puls_2006a}
{Puls}, J., {Markova}, N., {Scuderi}, S., {et~al.} 2006, \aap, 454, 625

\bibitem[{{Rahoui} {et~al.}(2011){Rahoui}, {Lee}, {Heinz}, {Hines},
  {Pottschmidt}, {Wilms}, \& {Grinberg}}]{Rahoui_2011a}
{Rahoui}, F., {Lee}, J.~C., {Heinz}, S., {et~al.} 2011, \apj, 736, 63

\bibitem[{{Reid} {et~al.}(2011){Reid}, {McClintock}, {Narayan}, {Gou},
  {Remillard}, \& {Orosz}}]{Reid_2011a}
{Reid}, M.~J., {McClintock}, J.~E., {Narayan}, R., {et~al.} 2011, \apj, 742, 83

\bibitem[{{Remillard} \& {Canizares}(1984)}]{Remillard_1984a}
{Remillard}, R.~A. \& {Canizares}, C.~R. 1984, \apj, 278, 761

\bibitem[{Rothschild {et~al.}(1998)Rothschild, Blanco, Gruber, Heindl,
  MacDonald, Marsden, Pelling, Wayne, \& Hink}]{Rothschild_1998a}
Rothschild, R.~E., Blanco, P.~R., Gruber, D.~E., {et~al.} 1998, ApJ, 496, 538

\bibitem[{{Sako} {et~al.}(1999){Sako}, {Liedahl}, {Kahn}, \&
  {Paerels}}]{Sako_1999a}
{Sako}, M., {Liedahl}, D.~A., {Kahn}, S.~M., \& {Paerels}, F. 1999, \apj, 525,
  921

\bibitem[{{Shaposhnikov} {et~al.}(2012){Shaposhnikov}, {Jahoda}, {Markwardt},
  {Swank}, \& {Strohmayer}}]{Shaposhnikov_2012a}
{Shaposhnikov}, N., {Jahoda}, K., {Markwardt}, C., {Swank}, J., \&
  {Strohmayer}, T. 2012, \apj, 757, 159

\bibitem[{{Shaposhnikov} \& {Titarchuk}(2006)}]{Shaposhnikov_2006a}
{Shaposhnikov}, N. \& {Titarchuk}, L. 2006, \apj, 643, 1098

\bibitem[{{Sowers} {et~al.}(1998){Sowers}, {Gies}, {Bagnuolo}, {Shafter},
  {Wiemker}, \& {Wiggs}}]{Sowers_1998a}
{Sowers}, J.~W., {Gies}, D.~R., {Bagnuolo}, W.~G., {et~al.} 1998, \apj, 506,
  424

\bibitem[{{Suchy} {et~al.}(2008){Suchy}, {Pottschmidt}, {Wilms}, {Kreykenbohm},
  {Sch{\"o}nherr}, {Kretschmar}, {McBride}, {Caballero}, {Rothschild}, \&
  {Grinberg}}]{Suchy_2008a}
{Suchy}, S., {Pottschmidt}, K., {Wilms}, J., {et~al.} 2008, \apj, 675, 1487

\bibitem[{{Sundqvist} \& {Owocki}(2013)}]{Sundqvist_2013a}
{Sundqvist}, J.~O. \& {Owocki}, S.~P. 2013, \mnras, 428, 1837

\bibitem[{{Sundqvist} {et~al.}(2012){Sundqvist}, {Owocki}, {Cohen},
  {Leutenegger}, \& {Townsend}}]{Sundqvist_2012b}
{Sundqvist}, J.~O., {Owocki}, S.~P., {Cohen}, D.~H., {Leutenegger}, M.~A., \&
  {Townsend}, R.~H.~D. 2012, \mnras, 420, 1553

\bibitem[{{Tarter} {et~al.}(1969){Tarter}, {Tucker}, \&
  {Salpeter}}]{Tarter_1969a}
{Tarter}, C.~B., {Tucker}, W.~H., \& {Salpeter}, E.~E. 1969, \apj, 156, 943

\bibitem[{{Verner} {et~al.}(1996){Verner}, {Ferland}, {Korista}, \&
  {Yakovlev}}]{Verner_1996a}
{Verner}, D.~A., {Ferland}, G.~J., {Korista}, K.~T., \& {Yakovlev}, D.~G. 1996,
  \apj, 465, 487

\bibitem[{{Vrtilek} {et~al.}(2008){Vrtilek}, {Boroson}, {Hunacek}, {Gies}, \&
  {Bolton}}]{Vrtilek_2008a}
{Vrtilek}, S.~D., {Boroson}, B.~S., {Hunacek}, A., {Gies}, D., \& {Bolton},
  C.~T. 2008, \apj, 678, 1248

\bibitem[{{Walborn}(1973)}]{Walborn_1973a}
{Walborn}, N.~R. 1973, \apjl, 179, L123

\bibitem[{{Wilms} {et~al.}(2000){Wilms}, {Allen}, \& {McCray}}]{Wilms_2000a}
{Wilms}, J., {Allen}, A., \& {McCray}, R. 2000, \apj, 542, 914

\bibitem[{{Wilms} {et~al.}(2006){Wilms}, {Nowak}, {Pottschmidt}, {Pooley}, \&
  {Fritz}}]{Wilms_2006a}
{Wilms}, J., {Nowak}, M.~A., {Pottschmidt}, K., {Pooley}, G.~G., \& {Fritz}, S.
  2006, \aap, 447, 245

\bibitem[{{Xiang} {et~al.}(2011){Xiang}, {Lee}, {Nowak}, \&
  {Wilms}}]{Xiang_2011a}
{Xiang}, J., {Lee}, J.~C., {Nowak}, M.~A., \& {Wilms}, J. 2011, \apj, 738, 78

\bibitem[{{Yamada} {et~al.}(2013){Yamada}, {Torii}, {Mineshige}, {Ueda},
  {Kubota}, {Gandhi}, {Done}, {Noda}, {Yoshikawa}, \&
  {Makishima}}]{Yamada_2013a}
{Yamada}, S., {Torii}, S., {Mineshige}, S., {et~al.} 2013, \apjl, 767, L35

\bibitem[{{Zdziarski} {et~al.}(2011){Zdziarski}, {Pooley}, \&
  {Skinner}}]{Zdziarski_2011a}
{Zdziarski}, A.~A., {Pooley}, G.~G., \& {Skinner}, G.~K. 2011, \mnras, 412,
  1985

\bibitem[{{Zi{\'o}{\l}kowski}(2014)}]{Ziolkowski_2014a}
{Zi{\'o}{\l}kowski}, J. 2014, \mnras, 440, L61

\end{thebibliography}

\appendix 
\section{Clumpy wind model simulations}

\begin{figure*}
\hfill\includegraphics[height=0.2\textheight]{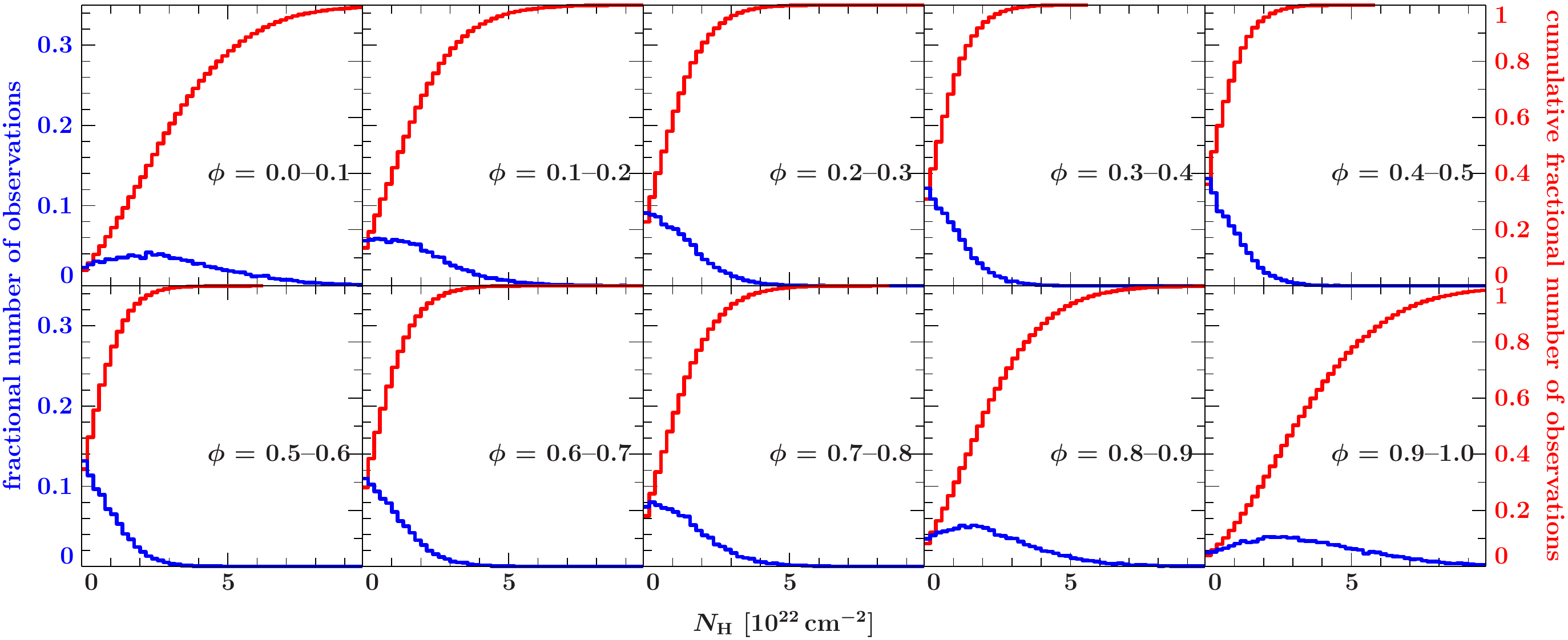}\hfill 
\includegraphics[height=0.2\textheight]{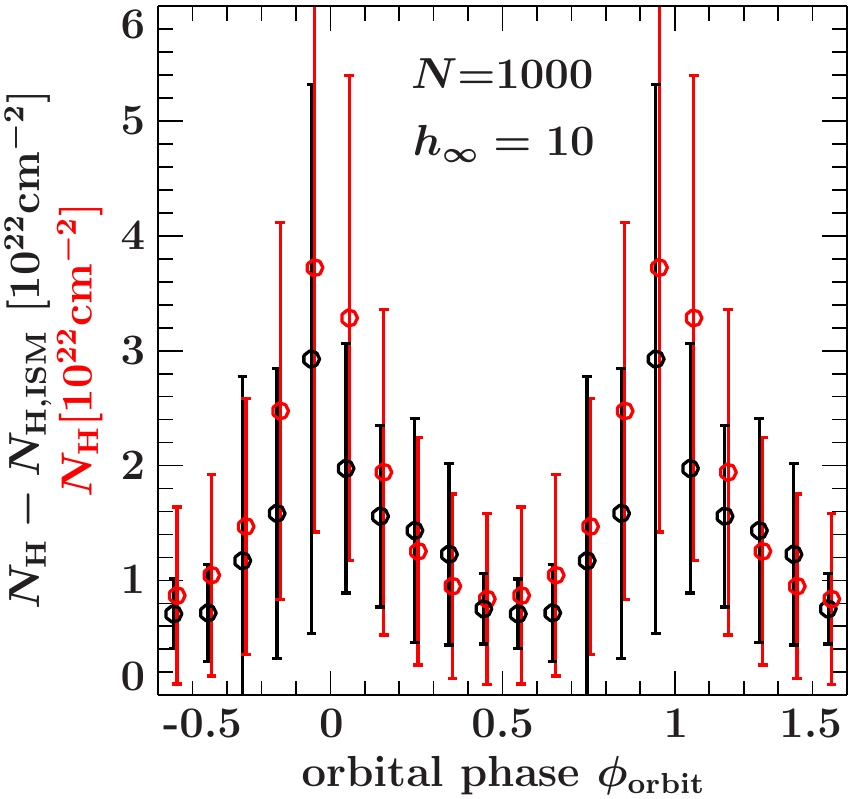}\hfill\\
\\
\hfill\includegraphics[height=0.2\textheight]{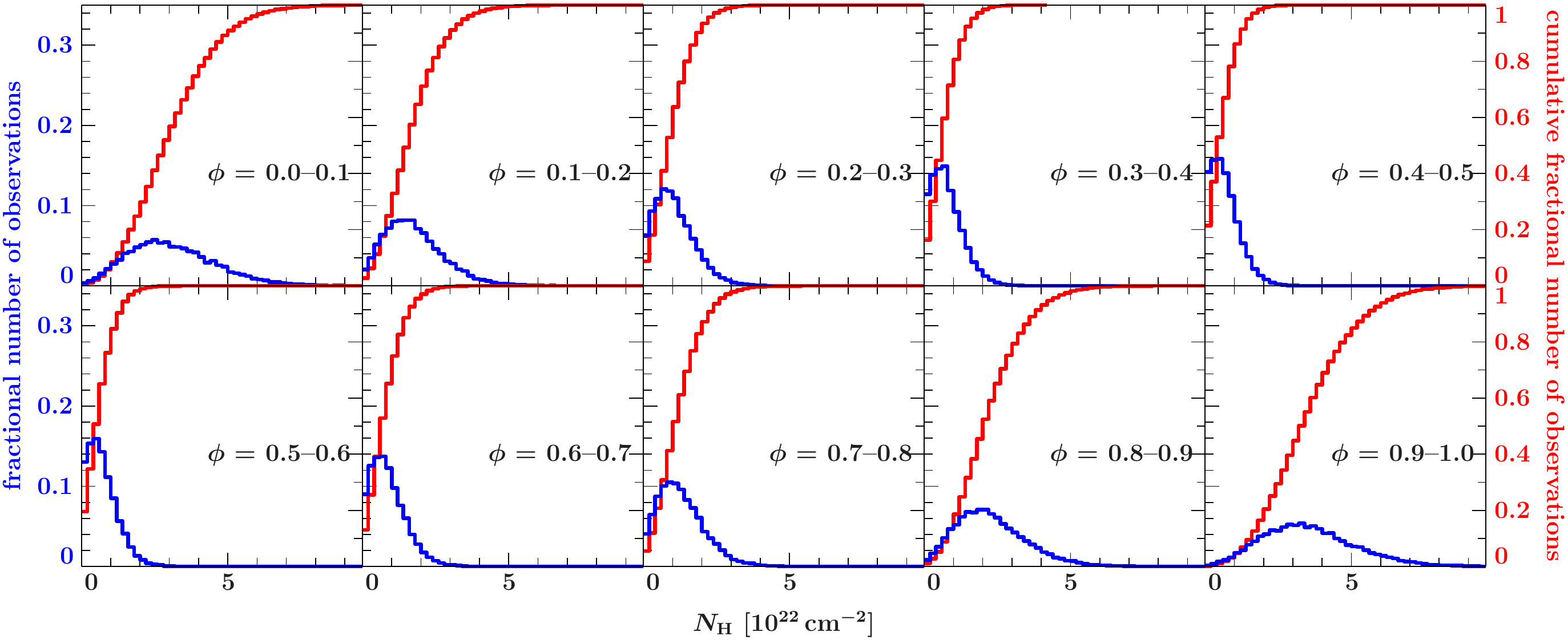}\hfill 
\includegraphics[height=0.2\textheight]{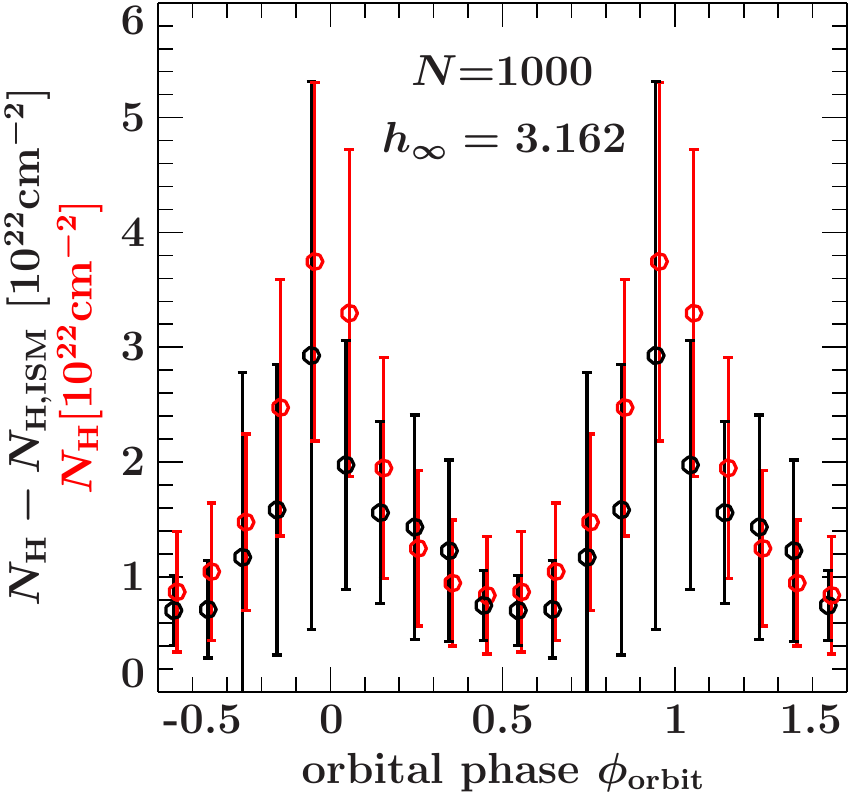}\hfill\\
\\
\hfill\includegraphics[height=0.2\textheight]{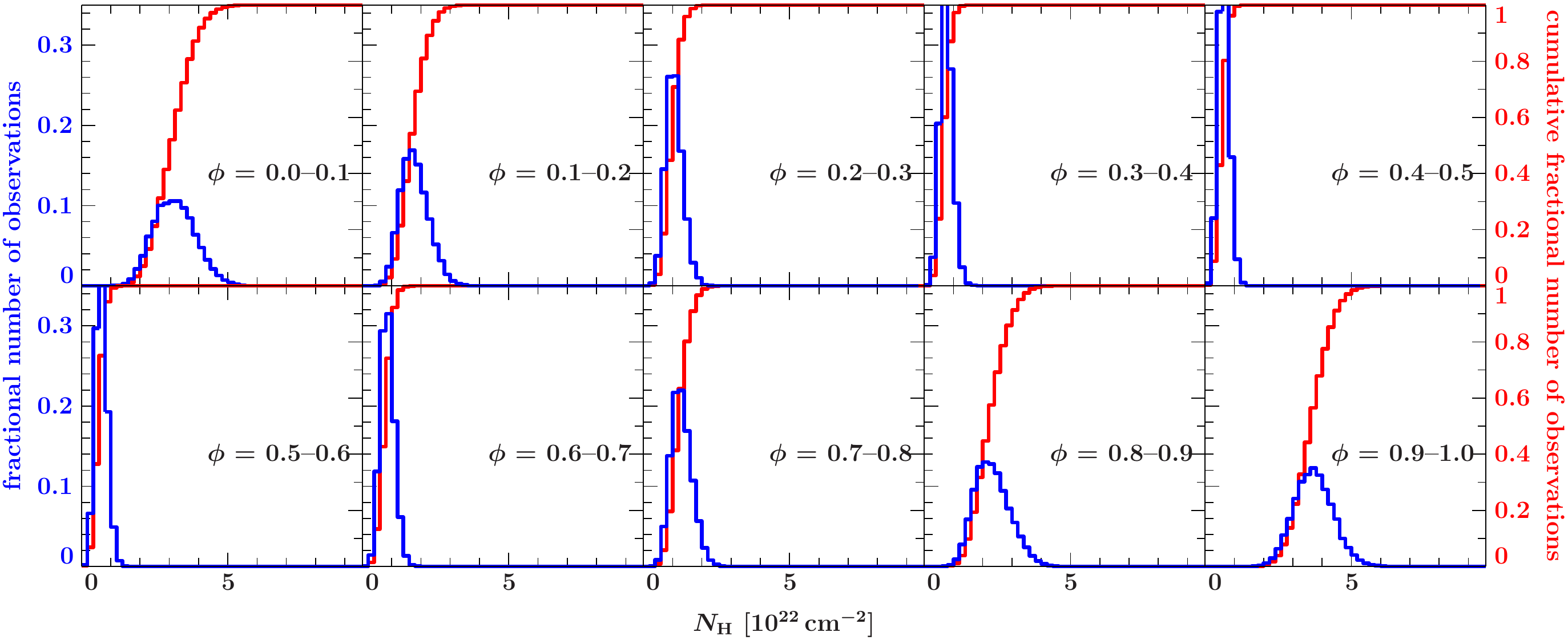}\hfill 
\includegraphics[height=0.2\textheight]{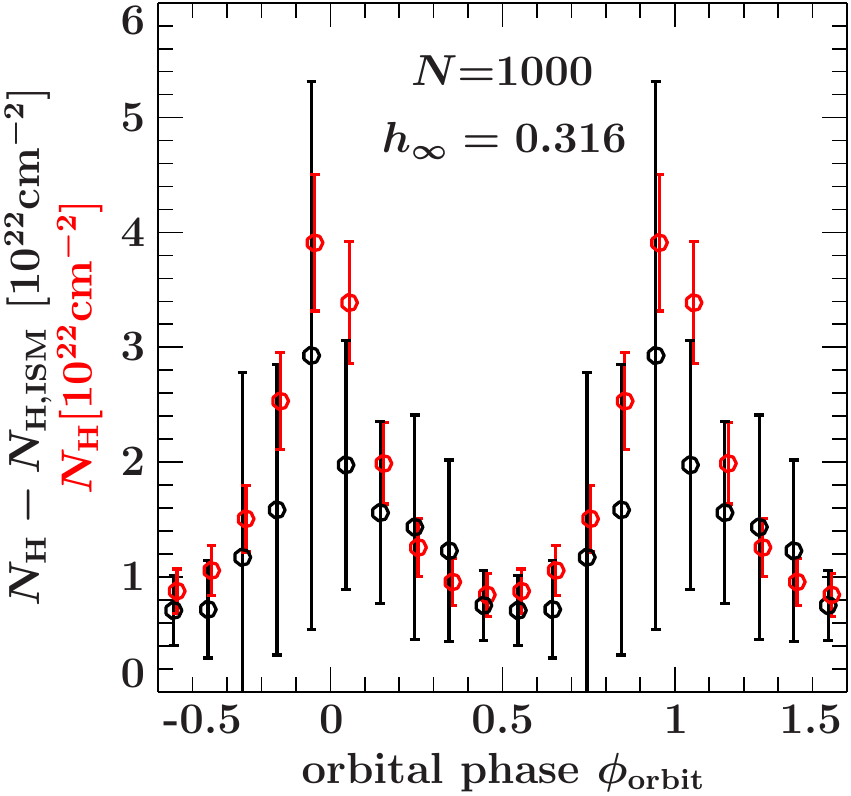}\hfill\\
\\
\hfill\includegraphics[height=0.2\textheight]{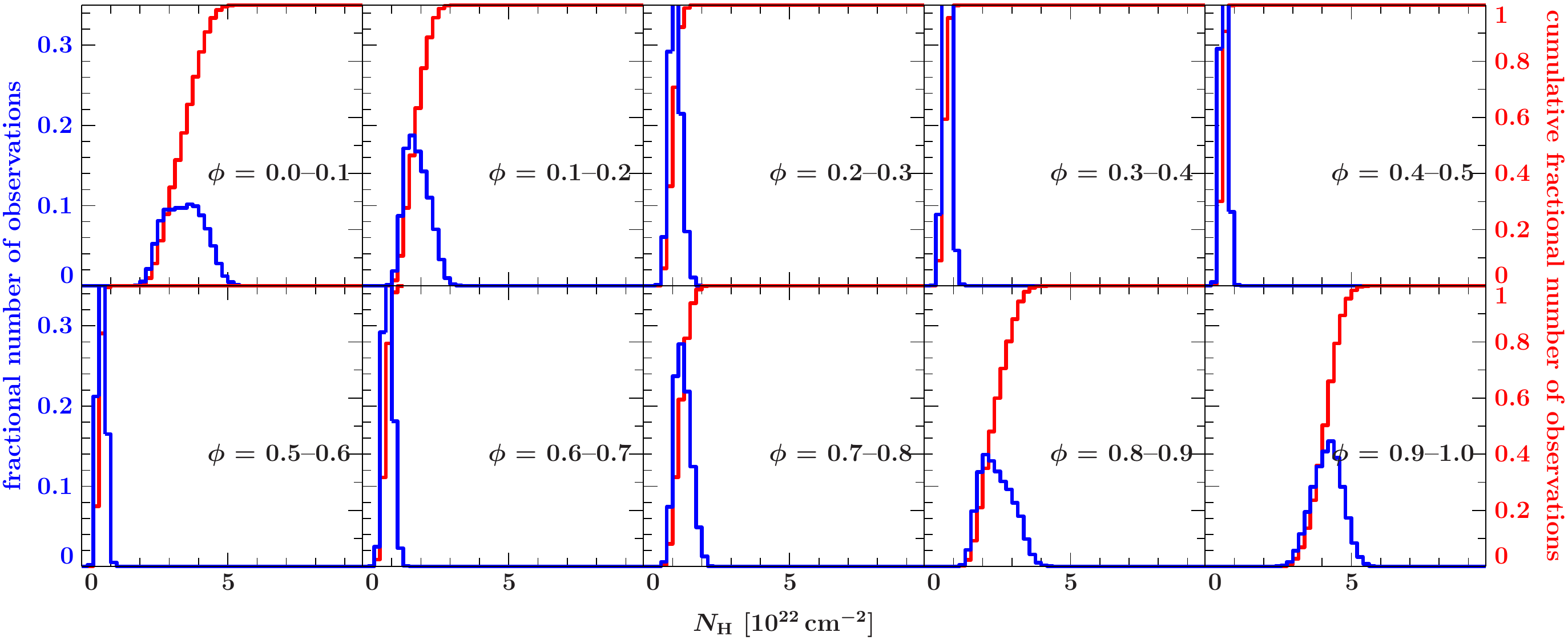}\hfill 
\includegraphics[height=0.2\textheight]{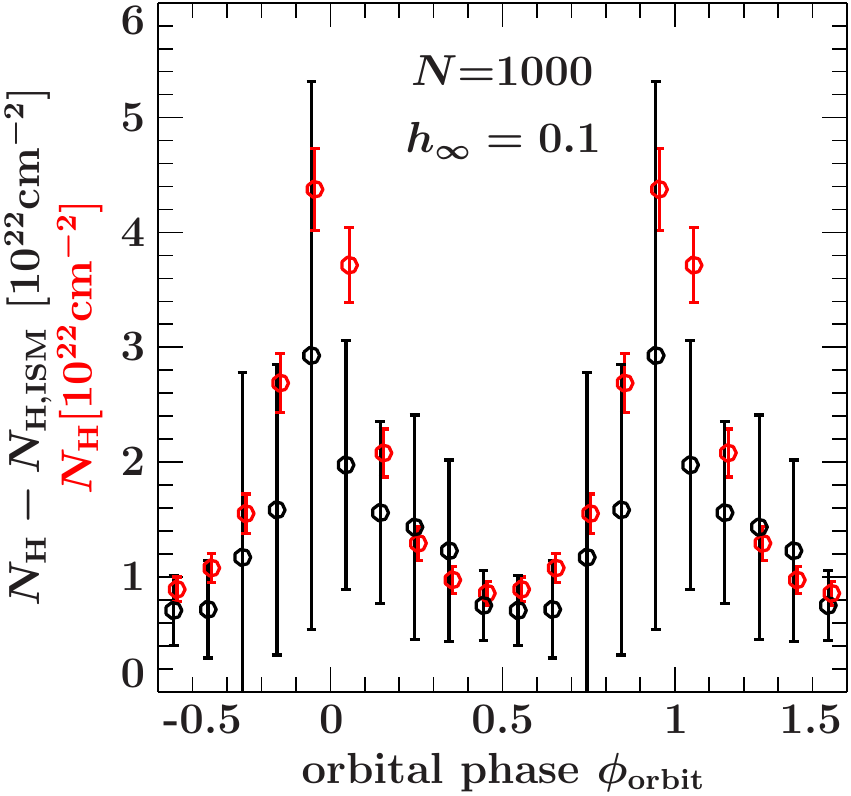}\hfill

\caption{Predictions of $N_\mathrm{H}$ variability assuming a clumpy
  wind model with the total number of clumps per flow time $N=1000$
  and varying terminal porosity length $h_\infty = 10$, 3.162, 0.316,
  and 0.1$R_*$ (see Sect.~\ref{sect:clumpy} for explanation of
  parameters and Fig.~\ref{fig:theory} for the case of $h_\infty =
  1$). \textsl{Left}: model histograms for the distribution of
  equivalent absorption column density $N_{\mathrm{H}}$.
  \textsl{Right}: average values (circles) and standard deviations
  (`error bars' on the average values) in a given orbital phase for
  the theoretical model (red) and values measured in the hard state of
  MJD 53900-55375 (black). Average measured values are shown as
  `$\mbox{average} - N_{\mathrm{H,ISM}}$' to account for the ISM
  absorption.
}\label{fig:theory_all}
\end{figure*}

\begin{figure*}
\includegraphics[width=\textwidth]{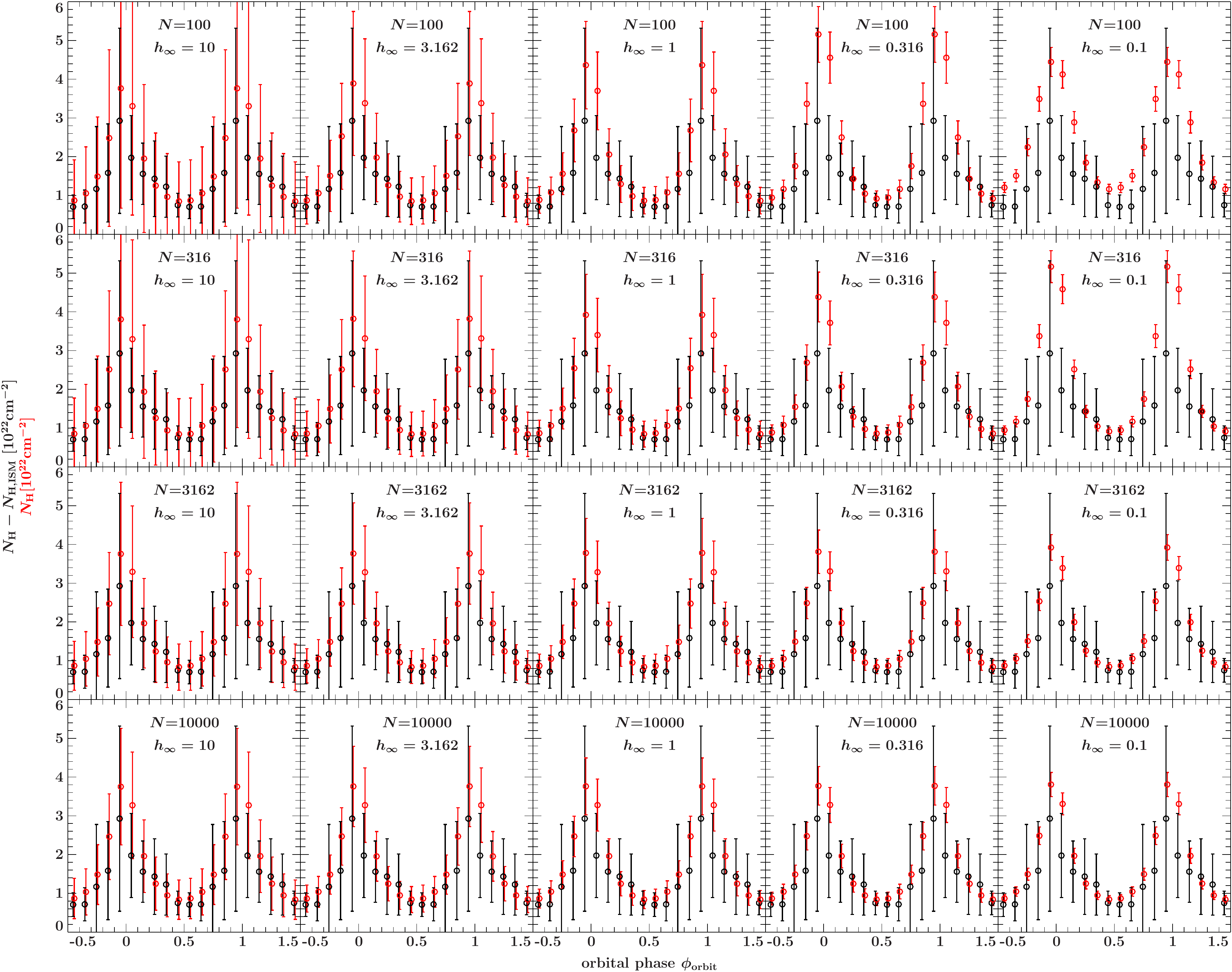}
\caption{Predictions of $N_\mathrm{H}$ variability assuming a clumpy
  wind model with the total number of clumps per flow time $N$ and
  varying terminal porosity length $h_\infty$.  Shown are average
  values (circles) and standard deviations (`error bars' on the
  average values) in a given orbital phase for the theoretical model
  (red) and values measured in the hard state of MJD 53900-55375
  (black). Average measured values are shown as `$\mbox{average} -
  N_{\mathrm{H,ISM}}$' to account for the ISM absorption. For the case
  of $N = 1000$ see Figs.~\ref{fig:theory} and ~\ref{fig:theory_all}.
}\label{fig:std_all}
\end{figure*}

\end{document}